\documentclass{ar-1col}

\usepackage[comma]{natbib}
\usepackage{url}
\usepackage{epsfig}
\usepackage{amsmath}
\usepackage{amssymb}
\usepackage{graphicx}
\usepackage{subcaption} 

\newcommand{\kms}{km~s\ensuremath{^{-1}}}
\newcommand{\msun}{$M_{\odot}$}
\newcommand{\nuvr}{NUV-$r$}
\newcommand{\mh}{$M_{H_2}$}
\newcommand{\mhi}{$M_{HI}$}
\newcommand{\mgas}{$M_{gas}$}
\newcommand{\mstar}{$M_{\ast}$}
\newcommand{\mdust}{$M_{\rm dust}$}
\newcommand{\must}{$\mu_{\ast}$}
\newcommand{\hi}{{H{\sc i}}}
\newcommand{\hmol}{H$_2$}

\newcommand{\tdep}{$t_{dep}({\rm H_2})$}
\newcommand{\tdepHI}{$t_{dep}$(\hi)}
\newcommand{\fgas}{$f_{H_2}$}
\newcommand{\fhi}{$f_{HI}$}
\newcommand{\xco}{$\alpha_{\rm CO}$}
\newcommand{\gd}{$\delta_{\rm GD}$}

\newcommand{\ms}{SFR-M$_{\ast}$}
\newcommand{\deltams}{$\Delta$(MS)}
\newcommand{\about}{$\sim$}
\newcommand{\omegahi}{$\Omega_{{\rm HI}}$}
\newcommand{\omegahmol}{$\Omega_{{\rm H_2}}$}

\newcommand{\nat}{nature}
\newcommand{\aj}{AJ}
\newcommand{\apj}{ApJ}
\newcommand{\apjs}{ApJS}
\newcommand{\aap}{AAP}
\newcommand{\aaps}{A\&AS}
\newcommand{\apjl}{ApJL}
\newcommand{\apss}{Ap\&SS}

\newcommand{\mnras}{MNRAS}

\newcommand{\araa}{Annu. Rev. Astron. Astrophys.}
\newcommand{\aapr}{AAPR} 

\newcommand\rmxaa{Rev. Mex. Astron. Astrofis.} 
\newcommand{\fcp}{Fund. Cosmic Phys.}

\newcommand{\pasa}{PASA}
\newcommand{\pasp}{PASP}

\jname{Xxxx. Xxx. Xxx. Xxx.}
\jvol{AA}
\jyear{YYYY}
\doi{10.1146/((please add article doi))}

\begin{document}

\markboth{Saintonge \& Catinella}{The Cold Interstellar Medium}

\title{The Cold Interstellar Medium of Galaxies in the Local Universe}

\author{Am\'elie Saintonge$^1$ and Barbara Catinella$^{2,3}$
\affil{$^1$Department of Physics \& Astronomy, University College London, London, UK, WC1E 6BT; email: a.saintonge@ucl.ac.uk}
\affil{$^2$International Centre for Radio Astronomy Research, The University of Western Australia, Crawley, WA, Australia}
\affil{$^3$ARC Centre of Excellence for All-Sky Astrophysics in 3 Dimensions (ASTRO3D)}}

\begin{abstract}
The cold interstellar medium (ISM) plays a central role in the galaxy evolution process. It is the reservoir that fuels galaxy growth via star formation, the repository of material formed by these stars, and a sensitive tracer of internal and external processes that affect entire galaxies. Consequently, significant efforts have gone into systematic surveys of the cold ISM of the galaxies in the local Universe. This review discusses the resulting network of scaling relations connecting the atomic and molecular gas masses of galaxies with their other global properties (stellar masses, morphologies, metallicities, star formation activity...), and their implications for our understanding of galaxy evolution. Key take-home messages are as follows: \\
$\bullet$ From a gas perspective, there are three main factors that determine the star formation rate of a galaxy: the total mass of its cold ISM, how much of that gas is molecular, and the rate at which any molecular gas is converted into stars. All three of these factors vary systematically across the local galaxy population. \\
$\bullet$ The shape and scatter of both the star formation main sequence and the mass-metallicity relation are deeply linked to the availability of atomic and molecular gas. \\
$\bullet$ Future progress will come from expanding our exploration of scaling relations into new parameter space (in particular the regime of dwarf galaxies), better connecting the cold ISM of large samples of galaxies with the environment that feeds them (the circumgalactic medium in particular), and understanding the impact of these large scales on the efficiency of the star formation process on molecular cloud scales.
\end{abstract}

\maketitle

\tableofcontents

\section{INTRODUCTION}

\subsection{The nature and importance of the cold interstellar medium}

A detailed understanding of the interstellar medium (ISM) is crucial for most astrophysical pursuits. The ISM is shaped by pressure, turbulence and magnetic fields, consumed by star formation and supermassive black holes, heated and enriched by stellar winds and supernova explosions, and replenished by accretion from the Intergalactic Medium (IGM) or from galactic fountains \citep{shapiro76}. Being at the crossroads of physical and chemical processes operating on such a range of scales, it is an incredibly rich source of information, but also a considerable challenge to model.

The ISM spans in excess of six orders of magnitude in temperature and density, with pressure equilibrium dictating that it breaks down into three distinct phases \citep{mckee77,cox05}: the hot ionised medium (HIM), the warm medium, including both a neutral (WNM) and an ionized (WIM) component, and the cold neutral medium (CNM). In a galaxy like the Milky Way, the HIM and WIM dominate by volume, but due to their low densities ($n\sim10^{-3}-10^{-1}~{\rm cm}^{-3}$), most of the mass is in the cooler and denser neutral phases \citep{draine11}. This cold ISM and its role in driving star formation and shaping galaxy evolution is the topic of this review.

\begin{figure}[h]
\includegraphics[width=4.5in]{ 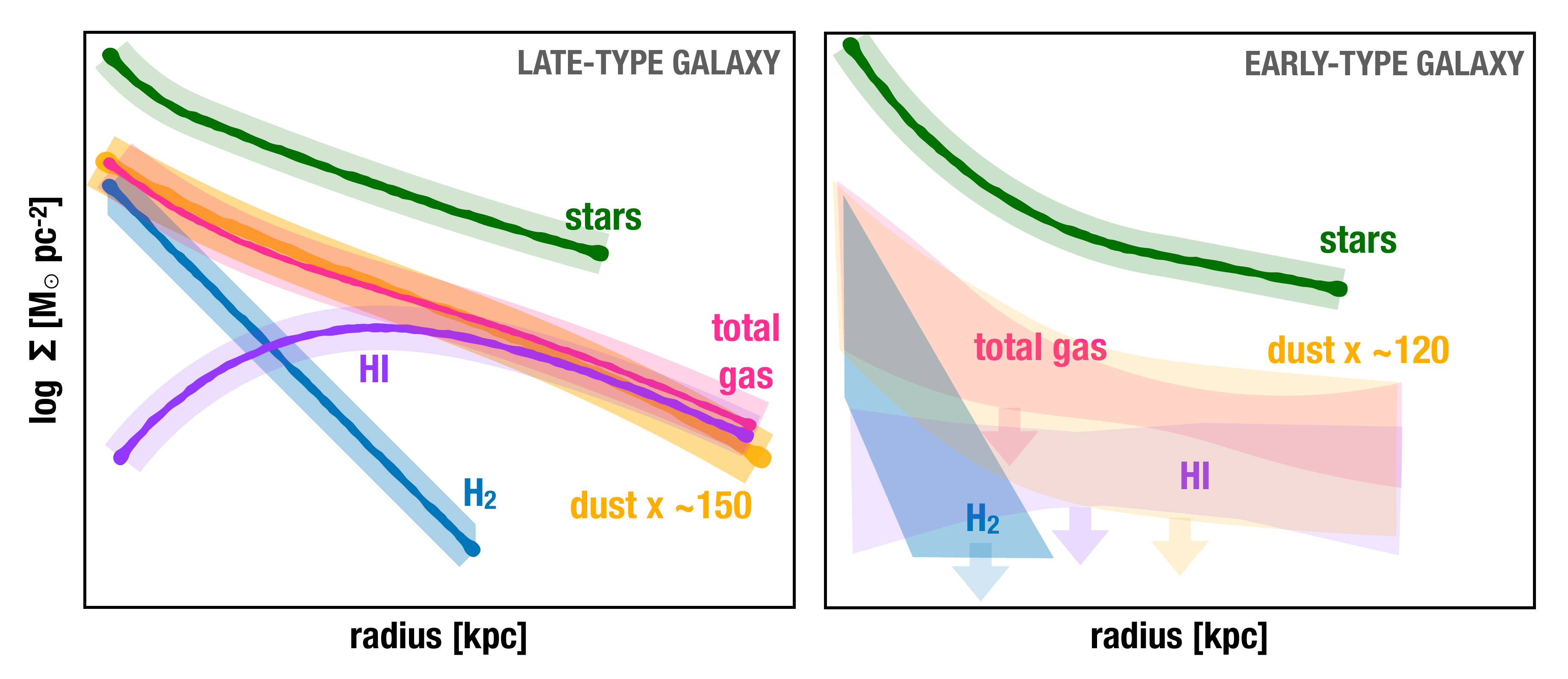}
\caption{Radial distribution of cold gas, dust and stars in ``typical" late- and early-type galaxies of the nearby Universe. The spiral galaxy sketch is inspired by \citet{casasola17} and \citet{bigielblitz15}. The elliptical model is based on ATLAS$^{\rm 3D}$ and Herschel Reference Survey results \citep{serra12,davis13,smith12}. The dust curves have been scaled up by a typical gas-to-dust ratio for late- and early-type galaxies \citep{delooze20,smith12}.}
\label{fig:gas_distribution}
\end{figure}

In a typical nearby star-forming late-type galaxy, the cold ISM accounts for $\sim10-20\%$ of the total baryonic mass (gas $+$ stars). Of that cold ISM, $\sim75\%$ by mass is atomic gas, mostly hydrogen (\hi), which is found in a warm ($\sim5000$~K), more diffuse ($\sim0.6~{\rm cm}^{-3}$) phase and in a cold, denser phase ($\sim100$~K, $\sim30~{\rm cm}^{-3}$), both contributing to the measured \hi\ emission at 21 cm \citep{draine11}.
The remaining $\sim25\%$ by mass is in molecular form, with temperatures of $\sim10-100$~K, and densities in the range of $\sim10^3-10^6~{\rm cm}^{-3}$. Dust, amounting to $\sim1\%$ of the gas mass, is mixed in with both the atomic and molecular components of the ISM, and therefore traces the total gas profile, as illustrated in Figure \ref{fig:gas_distribution}. Star formation occurs once surface densities of $\sim10$\msun pc$^{-2}$ are reached. Above this threshold, most of the ISM is molecular \citep{wong02,bigiel08}. The picture for early-type galaxies is rather different, with the cold ISM only accounting for $<3\%$ of the total baryonic mass. If present, any molecular gas tends to be centrally concentrated in a disc with a small scale length, but this gas can nonetheless have central surface densities similar to late-type galaxies \citep{davis13}. The atomic gas of early-type galaxies exhibits a huge range of possible morphologies \citep{serra12}.

\begin{marginnote}[]
\entry{Interstellar Medium (ISM)}{Interstellar matter within galaxies and filling the space between the stars.}
\entry{Intergalactic Medium (IGM)}{The diffuse hot plasma that fills the space between galaxies and traces the underlying dark matter cosmic web.}
\entry{Circumgalactic Medium (CGM)}{The hot ionised gas halo of galaxies located outside of the stellar component but within the virial radius. It forms the interface between the IGM and the ISM.}
\entry{Baryon cycle}{The lifecycle by which gas is accreted onto galaxies from the IGM/CGM, feeds star formation, and is returned to the external environment due to nuclear and stellar activity.}
\end{marginnote}

Significant progress has been made in the past decade to turn observations of the mass and distribution of the cold ISM into a framework of scaling relations that connect the gas with the internal and external properties of galaxies, as well as their redshift evolution \citep[e.g.][]{GASS1, COLDGASS1, boselli14_1, cicone17, tacconi10, tacconi20}. Observations of galaxies in the local Universe (loosely defined here as $z<0.05$) have been of great importance in painting this picture; only there do current facilities allow us to measure both the atomic and molecular gas across large samples of galaxies representative of the entire population. This review is centered around these observations, and the insights they bring into key questions in galaxy evolution. 

While the central role of gas in galaxy evolution has long been recognized \citep[e.g.][]{rees77,tinsley80}, these new gas scaling relations have triggered a resurgence of gas-centric galaxy evolution models \citep[e.g.][]{dekel09, bouche10, dave11b, lilly13}. These so-called ``equilibrium" or ``gas regulator" models rely on the balance between the various components of the baryon cycle: they equate the inflow rate of gas onto a galaxy (typically tied to the accretion rate of dark matter onto halos) to the rates at which this gas is locked up into stars via star formation and returned to the outside environment by outflows. Their success, especially given their simplicity, reinforces the view that gas is central to galaxy evolution.

Numerical simulations, on the other hand, strive to implement all of the physics required to produce the entire $z\sim0$ galaxy population, not just those galaxies that are in a state of equilibrium. Large volume simulations however lack the resolution to capture the physics of gas cooling and star formation, instead relying on ``subgrid" prescriptions. This includes for example the recipe for partitioning the cold ISM gas into \hi\ and \hmol, as well as prescriptions for star formation and stellar feedback. 
Interestingly, modern simulation suites all reproduce the general stellar and star formation properties of $z=0$ galaxies well, while relying on a wide range of such subgrid prescriptions. It is proving more challenging for these simulations to simultaneously reproduce observations of both atomic and molecular gas, but any discrepancies can be used to test and improve on the subgrid models for phenomena such as the \hi-to-\hmol\ transition and stellar feedback \citep[e.g.][]{lagos11,crain17,dave11b,dave20,diemer18}. Through these numerical efforts, global cold ISM measurements have the power to shed light on physics on much larger and smaller scales.

\subsection{Scope and objectives of this review}

This review is focused on how the total atomic and molecular gas masses of galaxies connect to their stellar mass, morphology, star formation activity, and other global properties, and how those dependencies fit in the broader galaxy evolution framework. Our main aims are to showcase how such global observations of the cold ISM can inform a range of key questions in the field of galaxy evolution (e.g. the nature and scatter of the galaxy main sequence, galaxy quenching, and the efficiency of the star formation process), by both being a direct agent in these processes, or by encoding information about phenomena that happen on much larger or smaller scales. 

Much about the link between the ISM and star formation has been learnt from carbon monoxide (CO) and \hi\ observations at $\sim$~kpc resolution of samples of up to $\sim50-100$, mostly spiral galaxies (e.g. HERACLES; \citet{leroy09} and EDGE; \citet{bolatto17} for molecular gas, THINGS; \citet{walter08} for \hi), and now even at resolution of tens of pc for CO studies \citep[e.g. PHANGS;][]{leroy21}. Results from such studies are touched upon as part of the discussions in Section \ref{sec:applications}, but will otherwise not be reviewed in detail; due to the unavoidable compromise between spatial resolution and sample size, they do not yet provide the full overview of the local galaxy population we are seeking here, but nonetheless are crucial in connecting global gas properties with relevant physical processes. Similarly, we do not review in detail the connection between the ISM and the larger scale gaseous environment, because an overall observational picture has yet to be fully established. This is an example of an area where there is significant scope for future discoveries with upcoming facilities, as discussed towards the end of this review.

\section{OBSERVING THE COLD ISM}
\label{sec:observations}

\subsection{Measuring the atomic ISM: the \hi\ 21-cm line}

Neutral atomic hydrogen (\hi) is the most common element in the Universe, and the main constituent of the ISM of nearby disk galaxies. \hi\ emits a spectral line at 21 cm, which was theoretically predicted by H. C. van de Hulst in 1944 and first detected in the Milky Way by \citet{ewen51}. The line arises from a magnetic dipole transition between the hyperfine structure levels of the hydrogen atom in its ground state. 
The splitting of the ground state into two energy levels is caused by the magnetic interaction between the electron and proton spins (with the parallel spin state having higher energy), and the frequency of the electron spin-flip transition between the two is one of the most precisely measured physical quantities, $\nu_{10}=1.420~405~751~786(30)$ GHz, corresponding to a wavelength of 21.106 cm. The transition probability is very small ($A_{10}=2.86888(7) \times 10^{-15} s^{-1}$), which translates into a radiative half-life for spontaneous de-excitation of $t_{1/2}\simeq 1/A_{10} \simeq 1.11 \times 10^7$~ yr. In most astrophysical situations, the relative population of the hyperfine levels is determined by collisions. 

Because of its long wavelength, \hi\ emission from external galaxies is spatially unresolved (except for the nearest systems) by single-dish telescopes, which have a spatial resolution (half-power beam width, HPBW) ranging from a few to tens of arcminutes. The construction of radio interferometers such as the Westerbork Synthesis Radio Telescope (WSRT) in the Netherlands (completed in 1970) and the Very Large Array (VLA) in New Mexico (completed in 1980), opened up the possibility to map the spatial distribution and velocity field of the \hi\ emission within galaxies. However, the gain in spatial resolution of radio interferometers comes at a cost as their smaller collecting areas (i.e., lower sensitivity), compared to large single-dish telescopes, means that \hi\ maps have become available almost exclusively for small samples of star-forming galaxies, generally within 100 Mpc from the Milky Way \citep[e.g.,][]{whisp,THINGS}.
Luckily, the next-generation \hi\ surveys on the Square Kilometre Array pathfinders (see Section~\ref{sec:Outlook}) have already begun to increase these samples. 

Even without spatial information, single-dish global \hi-line profiles provide three important parameters: accurate redshifts (the spectral resolution of \hi\ observations of galaxies is typically a few \kms\ or better), total \hi-line fluxes and velocity widths (a measure of the Doppler broadening). If a galaxy is unresolved by the radio telescope beam, its total \hi\ mass can be simply obtained from its measured \hi\ flux density integrated over the width of the line, assuming that the emission is optically thin\footnote{
{Much of the \hi\ mass of galaxies is in a warm phase with spin temperatures of several thousand K, which is optically thin. This assumption does not hold in denser regions of the ISM (e.g., \citealt{braun97,braun09}), but their filling factor is usually small. A correction for \hi\ self-absorption, significant for galaxies close to edge-on view, has sometimes been applied to the \hi\ mass (see e.g. appendix B in \citealt{hg84} and \citealt{springob05})}.
}, using the following expression:
\begin{equation}
    M_{\rm HI}[M_{\odot}] = \frac{2.356\times 10^5}{(1+z)^2} D_L^2
    \int S_\nu~dv
\label{eq_MHI}
\end{equation}
\noindent
where $D_L$ is the luminosity distance (in Mpc) to the galaxy at
redshift $z$, as measured from the \hi\ spectrum in the observed velocity frame, and the integral is the \hi-line flux density integrated over the line, with $S_\nu$ in units of Jy and $v$ in \kms.

The \hi\ spectral line proved to be a powerful tool to investigate not only the atomic gas reservoirs of galaxies and their kinematics, but also to determine extragalactic distances and peculiar motions of galaxies (via the Tully-Fisher relation; \citealt{tf77}) and map the large-scale structure of the nearby Universe, especially in low-density regions of superclusters (where late-type galaxies are abundant and radio redshifts easy to measure). Generally observed in emission, but also in absorption against continuum radio sources, the 21 cm line is not affected by interstellar dust and can therefore probe regions that are not accessible to optical telescopes.

\subsection{Measuring the molecular ISM: the CO molecule}

Most of the molecular ISM is composed of \hmol. The molecule was first detected in the Galactic ISM via Lyman resonance absorption bands \citep{carruthers70}, confirming theoretical predictions that in regions of the ISM with visual extinction in excess of one magnitude, most of the hydrogen is in molecular form  \citep{hollenbach71}. 

As for any other diatomic molecule, \hmol\ has electronic, vibrational and rotational energy levels. However, because \hmol\ is homonuclear and therefore perfectly symmetric, the vibrational and rotational states can only emit radiation through their quadrupole moment. These lines have low transition probabilities, but even more problematically, high excitation energies requiring either gas temperatures in excess of 500K or a strong UV radiation field, conditions that are much more the exception than the norm when it comes to the molecular ISM. This leaves observers in the unfortunate situation that the species accounting for most of the mass of the molecular ISM is almost completely invisible to them. 

However, over 200 different molecules have now been detected in the ISM of the Milky Way and other galaxies \citep{mcguire18}. Since each transition of each of these molecules has different excitation requirements, observers are armed with invaluable tools to probe the temperature and density of the molecular ISM. Most of these molecules however have very low abundances compared to \hmol, making them detectable in only the nearest or brightest of galaxies. For this reason, most of what we know about the molecular ISM of galaxies comes from observations of the second-most abundant molecule, carbon monoxide (CO). These observations are therefore the focus of this review.

The lowest energy transition of the $^{12}$CO molecule, at just $h\nu/k_B = 5.5$~K and with a critical density of $\sim3000$~cm$^{-3}$, is the $J=1\rightarrow0$ rotational transition (hereafter referred to as the CO(1-0) emission line\footnote{In this review, CO always refers to $^{12}$C$^{16}$O. Other species such as $^{13}$C$^{16}$O and $^{12}$C$^{18}$O are also present in the ISM, but less widely observed as significantly fainter due to low abundances. The flip side is that while $^{12}$C$^{16}$O is generally optically thick, the isotopologues are optically thin and therefore more sensitive to the gas density.}). These properties are conveniently very well matched to the typical conditions in the diffuse molecular phase of the ISM and in molecular clouds, which have temperatures of $\sim10-50$~K and densities of $\sim10^2-10^5$~cm$^{-3}$ \citep{draine11}. In addition, the wavelength of the CO(1-0) emission line, at 2.6mm, falls in a relatively transparent atmospheric window, making it accessible to ground based single-dish and interferometric millimetre telescopes. 

Most of what we know of the molecular ISM in galaxies of the nearby Universe  therefore comes from observations of the CO(1-0) line. The observed CO(1-0) line flux is converted into a luminosity $L^{\prime}_{\rm CO}$ \citep[see e.g][]{solomon97}, which is then itself turned into a total molecular gas mass using the CO-to-H$_2$ conversion factor, \xco: 
\begin{equation}
    M_{\rm H_2}[M_{\odot}] = \alpha_{\rm CO} L^{\prime}_{\rm CO(1-0)} = \alpha_{\rm CO} \left [ 3.25\times10^7 S_{\rm CO}~\nu_{obs}^{-2}~D_L^2~(1+z)^{-3} \right ], 
    \label{eq:MH2}
\end{equation}
where $L^{\prime}_{\rm CO(1-0)}$ is the line luminosity in units of K~\kms~pc$^2$, $S_{\rm CO}$ the integrated line flux in Jy~\kms, $\nu_{obs}$ the observed frequency of the CO(1-0) line in GHz and $D_L$ the luminosity distance in Mpc. For ISM conditions like the Milky Way's, the conversion factor is generally taken to be \xco$=3.2$~\msun~(K~\kms~pc$^2)^{-1}$ (this value is not inclusive of the correction to account for the contribution of helium and heavier elements; see Section \ref{sec:Helium}). There are comprehensive reviews of the use of CO line fluxes for measuring the molecular gas mass of galaxies in \citet{bolatto13} and \citet{tacconi20}, including a detailed account of why CO(1-0) ``works" as a molecular gas tracer despite being generally optically thick, how \xco\ is measured and how it may deviate from the Galactic value in certain environments. Indeed, a significant challenge with using CO(1-0) as our molecular gas tracer of choice is that \xco\ is actually a conversion \emph{function}, varying with the conditions of the ISM, in particular the dust abundance and the radiation field environment \citep[e.g.][]{israel97,accurso17b}.

\subsection{Measuring the cold ISM: dust-based tracers}
\label{sec:indirecttracers} 
It is also possible to infer the cold gas mass of galaxies from their dust contents, a method that has attracted significant interest in recent years with observatories like {\it Herschel} providing a wealth of far-infrared (FIR) data. Assuming that dust is well mixed with both the cold atomic and molecular phases of the ISM, the total gas mass of a galaxy can be inferred from the dust mass, \mdust, and a gas-to-dust ratio, \gd: 
\begin{equation}
    M_{gas} = M_{\rm H\sc{I}} + M_{\rm H_2} = M_{\rm dust} \times \delta_{\rm GD}. 
    \label{eq:MH2GD}
\end{equation}

The total dust masses required to apply this method are best inferred from observations that probe the peak of the FIR spectral energy distribution (100-200 $\mu$m) or even longer wavelengths on the Rayleigh-Jeans tail, where emission from cold dust dominates. They can be derived from the photometry using either empirical \citep[e.g.][]{scoville14,gordon14,lamperti19} or physically-motivated models \citep[e.g.][]{draineli07}. A detailed description of dust mass measurements techniques in nearby galaxies can be found in \citet{gallianoARAA}. The second ingredient in Equation \ref{eq:MH2GD} is the gas-to-dust mass ratio, \gd. The value of \gd\ is linked to metallicity ($Z$), and in many applications a simple scaling with metallicity of \gd$\propto Z^{-1}$ is assumed \citep{issa90,leroy11}. Studies of the \gd($Z$) relation however reveal important scatter across the local galaxy population, and a possible steepening of the relation at lower metallicity \citep{remyruyer14,devis17}. For nearby galaxies, the method summarised in Eq. \ref{eq:MH2GD} is best used to estimate total gas masses, $M_{gas}$, or to estimate \mh\ when resolved maps of dust and \hi\ are available. This allows for a pixel-by-pixel estimation of \mh, after subtraction of the \mhi\ contribution to the total gas mass. When resolved maps of the CO distribution are also available, the technique is also often used to estimate \xco\ by combining Equations \ref{eq:MH2} and \ref{eq:MH2GD} \citep[e.g.][]{israel97,sandstrom13}. 

An alternative route is to use the extinction caused by the dust in the UV/optical \citep{bohlin78}. So far this has been explored mostly for molecular gas mass estimations, using the empirical correlation between the Balmer decrement (or total $V$-band attenuation, $A_V$) and \mh\ \citep[e.g.][]{concas19,yesuf19}. More accurate results can be obtained from modeling that takes into account variations in the gas-to-dust and dust-to-metal ratios, possible deviations from the Case B recombination scenario, and various dust geometries  \citep[e.g.][]{brinchmann13,piotrowska20}. When resolved spectroscopic observations are available, a correlation is found between $A_V$ estimated from the Balmer decrement and both dust and \hmol\ surface density maps, but with significant scatter and galaxy-to-galaxy variations \citep{kreckel13,barrera20}. Investigations into the nature of any parameters that may be responsible for driving this scatter, including the contribution of \hi, are necessary to enable accurate gas masses to be inferred.

Care must however be taken in interpreting gas masses inferred from dust tracers, with specific calibrations performing better at reproducing either the atomic, molecular, or total cold gas masses of nearby galaxies \citep[e.g.][]{groves15}. For example, \citet{janowiecki18} find that compared to other {\it Herschel} bands and more complex dust mass methods, the luminosity at 500$\mu$m ($L_{500}$) best predicts \mgas\ (0.15 dex scatter), while $L_{160}$ most directly estimates \mh\ (0.2 dex scatter). For relatively massive (\mstar $\gtrsim 10^{10}$\msun) and metal-rich galaxies,  FIR-predicted \mh\ values are in good agreement with CO measurements  \citep[e.g.][]{genzel15}, with intrinsic scatter of $<0.2$dex.  However, the applicability of the method, whether using dust in emission or absorption, needs to be further investigated for quiescent galaxies, in the lower mass / metallicity regime, in the presence of an AGN, and as a function of physical scale.

\subsection{The contribution of helium and metals} 
\label{sec:Helium}
In the ISM, helium and heavier elements are mixed in with the hydrogen (be it atomic or molecular) and therefore contribute to the total mass. To account for this, it is standard in the molecular gas community to apply a 36\% upward correction on the molecular hydrogen mass obtained from any of the above methods. For example, this is done by using a Galactic \xco\ factor of 4.35 \msun [K \kms\ pc$^2$]$^{-1}$, as opposed to the value of 3.2 \msun [K \kms\ pc$^2$]$^{-1}$  that would be needed to infer a ``pure" molecular hydrogen mass \citep{bolatto13}. This leads to the somewhat confusing situation where `` \mh" is used to represent the total mass in the molecular ISM, when a symbol of ``$M_{\rm mol gas}$" might be more appropriate. In contrast, the standard practice in the \hi\ community is to use \mhi\ calculated using Equation~\ref{eq_MHI}, accounting only for the hydrogen. The 36\% correction is sometimes applied on a case-by-base basis, when a total mass for the atomic medium is sought. 
To ensure homogeneity and avoid confusion, we adopt the following convention across this review: 
\begin{itemize}
    \item Quantities with the subscripts ``\hi" refer to atomic hydrogen, without heavier element correction, as standard in the \hi\ 21cm literature. 
    \item Quantities with the subscripts ``\hmol" refer to molecular hydrogen only, without helium and heavier elements. This is a departure from common practice in the field, but makes for a more homogeneous presentation and easier comparison with the \hi\ results. All quantities with the ``\hmol" subscript can be scaled up by 36\% for comparison with results in the literature when appropriate. 
    \item Quantities with the subscripts ``mol" refer to the total molecular gas, including the contribution of helium and metals. 
    \item Quantities with the subscripts ``gas" refer to the total cold ISM, therefore including \hi, \hmol, as well as helium and all the heavier elements. 
\end{itemize}

\subsection{Extragalactic cold ISM surveys} 
\label{sec:HIsurveys}
Since the first extragalactic detections of \hi\ in 1953 \citep{kerr53,kerr54} and CO two decades later \citep{rickard75,solomon75,combes77}, huge effort has gone into measuring the global amount, and more recently the detailed spatial distribution and kinematics, of these species in galaxies in the local Universe and beyond. Interestingly, the scope of these surveys has shifted from mapping the large-scale structure with the \hi-line (as 21 cm redshifts of star-forming galaxies can be obtained very quickly with single-dish telescopes), to investigations of the statistical properties of the cold gas content in large galaxy samples on one hand, and of the physics connecting cold gas and star formation in individual objects or small samples on the other. We focus here on \hi\ and \hmol\ surveys that are particularly relevant to the study of global scaling relations, and refer the reader to \citet{giovanelli15,wallaby} and references therein for more comprehensive historical overviews.

Most of the global cold gas measurements available today come from 
blind \hi-selected surveys. The largest and most impactful of these, the HI Parkes All Sky Survey \citep[HIPASS;][]{HIPASS} and the Arecibo\footnote{
The 305-meter diameter Arecibo radio telescope tragically collapsed on December 1st, 2020, after nearly 60 years of operation that have led to numerous ground-breaking discoveries. Its contributions had a tremendous impact for \hi\ astronomy, and in particular for the topic of this review. 
} 
Legacy Fast ALFA Survey \citep[ALFALFA;][]{alfalfa1}, catalogued over 5000 galaxies out to $z<0.04$ \citep{HIPASS_cat,HIPASS_north} and \about 31,500 sources out to $z<0.06$ \citep{haynes18}, respectively. These surveys have been invaluable to enable statistical studies such as \hi\ mass and velocity functions, as well as characterise the gas-rich galaxy population. The main drawbacks are their limited spatial resolutions (with a HPBW of \about 15.5 arcmin for HIPASS and \about 3.5 arcmin for ALFALFA at the rest frequency of the \hi\ line) and sensitivities, which lead to a bias towards the most gas-rich systems within their volumes. At the same time, targeted \hi\ studies have focused on galaxy populations and/or environments that were not well probed by large-area blind \hi\ surveys, such as early-type systems \citep[e.g.,][]{knapp85,morganti06,serra12} and galaxies in different environments, from isolated to nearby groups and clusters \citep[e.g.,][]{verdes01,kilborn09,viva}. 

The samples with molecular gas measurements are considerably smaller, due to the lack of blind CO-selected surveys at $z\sim 0$. Particularly remarkable was the FCRAO Extragalactic CO Survey \citep{young95}, which measured CO(1-0) in 300 nearby galaxies and remained the largest comprehensive study of the molecular ISM in the local Universe for 25 years. Because of a strong correlation between infrared and CO luminosities \citep[e.g.][]{sanders91}, a significant part of early extragalactic CO observations was done by  targeting luminous infrared galaxies \citep[e.g.][]{radford91,solomon97}. Recognizing this bias toward ``exceptional" galaxies (e.g. starbursts and interacting systems), efforts were made to target a broader range of galaxies. Samples of galaxies with integrated molecular line observations grew to include ``normal" non-interacting spiral galaxies \citep{braine93,sage93}, cluster spiral galaxies \citep{kenney88,boselli97}, early-type galaxies \citep{combes07,krips10,young11}, galaxies with active nuclei \citep{helfer93,sakamoto99,garcia03}, and isolated galaxies \citep{lisenfeld11}. 

The past decade has seen a shift from targeting specific types of galaxies, to surveying samples representative of the entire galaxy population in both \hi\ and CO lines. This became possible also thanks to the availability of large multi-wavelength surveys such as SDSS \citep{SDSS}, 2MASS \citep{2MASS}, and GALEX \citep{GALEX}, which measured stellar masses, star-formation rates and a plethora of other properties that could be used to define representative samples for \hi\ and CO follow up. The largest single cold gas surveys in the nearby Universe are the extended GALEX Arecibo SDSS Survey \citep[xGASS;][]{GASS1,GASS6, GASS8,xGASS} and the extended CO Legacy Database for GASS \citep[xCOLD GASS;][]{COLDGASS1,COLDGASS2,saintonge17}, which obtained \hi\ and CO observations for 1179 and 532 galaxies, respectively, at $0.02<z<0.05$. The objects were selected from SDSS and are representative of the full galaxy population with \mstar$>10^{9}$\msun. The strength of these surveys comes from their sample size, the homogeneity of all the data products, and the depth of the observations. As will be shown in Section \ref{sec:scalingrelations}, the latter is of particular importance as it makes it possible to infer the cold gas properties of even the most gas-poor sub-populations, via either spectral stacking or the setting of stringent upper limits. Other notable global cold gas surveys include the \hi\ census for the REsolved Spectroscopy Of a Local VolumE \citep[RESOLVE;][]{stark16} survey, designed to be baryonic mass limited;  the Herschel Reference Survey \citep[HRS;][]{boselli10}, which includes 322 galaxies (225 of which have CO observations and 315 with \hi\ data from the literature) that span the full range of environments, from low density regions to the core of the Virgo cluster; the ALLSMOG survey \citep{bothwell14,cicone17}, which observed CO(2-1) in 88 galaxies in the mass range $10^{8.5}<$\mstar$<10^{10}$\msun, increasing the coverage of parameter space towards lower mass galaxies. These surveys have been key in making cold gas a mainstream element in the multi-wavelength toolbox of extragalactic observers.

\section{GLOBAL STATISTICAL PROPERTIES}

Before delving into the details of the multi-parameter relations connecting the cold ISM masses of galaxies with their star formation activity and other global properties, we start by asking much simpler questions: how much molecular and atomic gas is there in the local Universe, and where is that gas predominantly located?

In an attempt to resolve the ``missing satellite problem", one of the science goals of the large-area blind \hi\ surveys described in Section \ref{sec:HIsurveys} was to search for ``dark galaxies". These objects would have significant \hi\ gas reservoirs, yet be optically-dark. The existence of these galaxies would not only help to bridge the gap between the number of dark matter halos and the optically-bright observed galaxies, but also represent a challenge for star formation models; how can galaxies have so much gas yet be unable to form stars? It would also indicate that a complete census of the cold ISM contents of the local Universe would have to take into account all this gas not associated with optically-detected galaxies. Most of the candidate dark galaxies have turned out to be intergalactic gas clouds, the result of galactic interactions or environmental processes \citep{cannon15}; for example one of the first reported dark galaxies, VIRGOHI21 \citep{minchin05}, was later found to be part of a large \hi\ tidal feature stripped from NGC4254 by its fall into the Virgo cluster \citep{haynes07}. Given that massive clusters where such events can take place are rare, the contribution of such interstellar gas clouds to the total \hi\ mass budget in the local Universe is negligible. 

\begin{marginnote}[]
\entry{Dark galaxy}{A class of galaxies theorised to have large \hi\ gas reservoirs yet no detectable stellar populations.}
\entry{Missing satellite problem}{The under-abundance of observed low mass galaxies compared with the number of equivalent dark matter halos predicted by cosmological simulations.}
\end{marginnote}

While we cannot discount the existence of a population of very low mass gas-rich but optically-faint objects below the detection limit of current blind \hi-selected surveys, these results nonetheless suggest that the vast majority of the cold atomic gas in the nearby Universe is located in galaxies with well-established stellar populations. We do not have similar blind molecular gas surveys at $z\sim0$, but it is reasonable to expect that the above statement is even more true for molecular gas, given the minimum pressure and density requirements for molecules to form out of atomic gas. 

\subsection{Cold gas mass functions} 

We can now look at the distribution of this gas by drawing the atomic and molecular gas mass functions. Having established that the vast majority of the cold gas resides in the galaxy population traced by optical surveys, these mass functions can be built using either blind or targeted (mostly optically-selected) samples; once completeness effects have been taken into consideration they should produce similar results \citep{fletcher21}. For samples that are flux-limited, mass functions are most often built using the ``$1/V_{max}$" method \citep{schmidt68}. This method accounts for Malmquist  bias by weighting each galaxy by the inverse of the volume in which galaxies of that specific brightness are detectable. Variations on this method exist to take into account and correct for large scale structure in the observed volume \citep{efstathiou88}. 

\begin{figure}[h]
\includegraphics[width=5in]{ 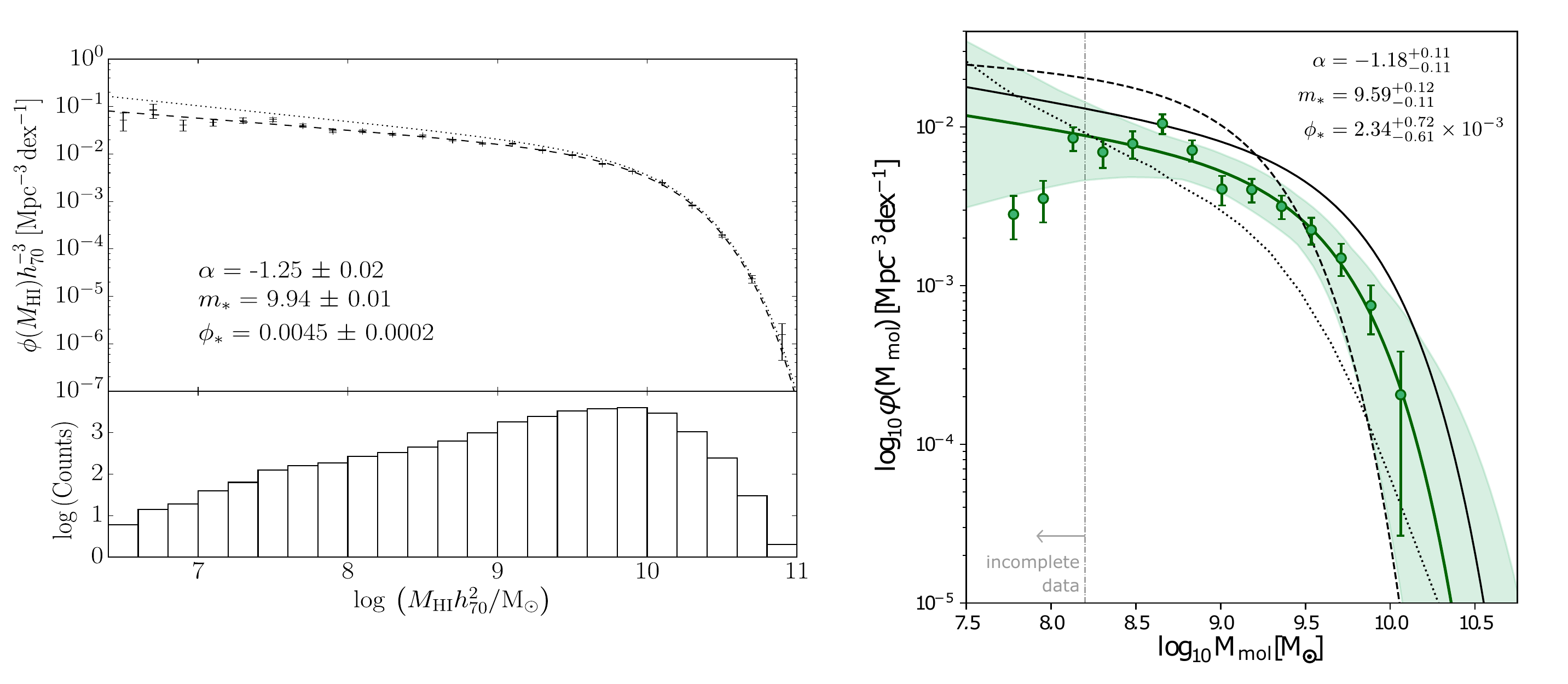}
\caption{{\bf Left}: The \hi\ mass function from the ALFALFA survey, and {\bf right}, the molecular gas mass function from xCOLDGASS. Figures adapted with permission from \citet{jones18} and \citet{fletcher21}. The parameters of the best-fitting Schechter functions (Eq. \ref{eq:schechter}) are given in each case (where $m_{\ast} = \log M^{\ast}$, the characteristic mass).}
\label{fig:HIH2MF}
\end{figure}

Both the \hi\ and \hmol\ mass functions (as well as the CO luminosity function) are well represented by a single \citet{schechter76} function of the form: 
\begin{equation}
    \phi(M) = \ln(10) \phi_{\ast} \left( \frac{M_{g}}{M^{\ast}} \right)^{\alpha+1}~e^{-M_{g}/M^{\ast}}, 
    \label{eq:schechter}
\end{equation}
where $M_{g}$ would be either the \hi\ or \hmol\ mass (or their sum if interested in a total cold gas mass function). In this equation, $M^{\ast}$ is the characteristic mass at which the function transitions from a power law with slope $\alpha$ to an exponential function, and $\phi_{\ast}$ is a normalisation factor. A Schechter function fit to the \hi\ and \hmol\ mass functions is shown in Figure~\ref{fig:HIH2MF}. 

Most of the blind \hi-selected surveys described in Sec. \ref{sec:HIsurveys} have been used to determine the \hi\ mass function. Differences between their results can be explained by the specific volume and depth of the observations. For example, shallow surveys tend to underestimate $M^{\ast}$ and the overall abundance of massive \hi\ systems (being rare, they are easily missed if the volume observed is too small; \citealt{martin10}). The footprint of the observations with respect to the large scale structure of the nearby Universe also has a direct impact on the \hi\ mass function; \citet{jones18} report a steeper low-mass slope $\alpha$ in the region around the Virgo cluster and a flatter slope in under-dense environments. This may be explained by a population of gas-rich galaxies in the filaments around clusters. Results for the shape of the \hi\ mass function in specific galaxy groups can vary significantly, possibly highlighting the importance of the larger scale environment into which these groups are found \citep{stierwalt09,pisano11,westmeier17,jones18}. 

Measurements of the \hmol\ mass function are far fewer between due to the challenges in assembling the necessary samples. \citet{keres03} built a CO luminosity function and \hmol\ mass function from the FCRAO Extragalactic CO Survey \citep{young95}, for both a far infrared flux-limited sub-sample, and an optical B-band–selected sub-sample. To avoid the possible biases associated with the IR- or optical-flux selection, robust \hmol\ mass functions have recently been computed using the large HRS and xCOLD GASS mass-selected samples \citep{andreani20,fletcher21}. In addition to the sample selection, the choice of the CO-to-\hmol\ conversion function also has a significant impact on the \hmol\ mass function and the best-fit Schechter function parameters  \citep{obreschkow09,fletcher21}.

\subsection{How much \hi\ and \hmol\ is there in the nearby Universe?} 

By integrating the mass functions, we obtain a measurement of \omegahi\ and \omegahmol, the cosmic mass densities of \hi\ and \hmol\ in the nearby Universe. Amongst many other measurements of \omegahi\ \citep[e.g.][]{zwaan05}, the ALFALFA survey estimates that \omegahi $=(3.9 \pm 0.1 \pm 0.6) \times 10^{-4}$ \citep{jones18}, while for molecular gas \citet{fletcher21} estimate $\Omega_{mol}=(7.62 \pm 0.47) \times 10^{-5}$ from the xCOLD GASS sample, and \citet{andreani20} report a value of $(1.1 \pm 0.4) \times 10^{-4}$ for the HRS sample. Once accounting for the presence of Helium and heavier elements by multiplying \omegahi\ by a factor of 1.36 (already included in $\Omega_{mol}$), the total density parameter for the ISM in the nearby Universe is found to be $\Omega_{ISM} = 0.00061$. Accounting for dust would increase this number by $\sim$1\% given standard dust-to-gas ratios measured in the local Universe \citep[e.g.][]{remyruyer14}. This corresponds to \about35\% of the total mass density of stars, or \about1.3\% of all the baryons\footnote{These values assume that $\Omega_{\ast} = 0.0017$ \citep{baldry12} and $\Omega_b = 0.0456$ \citep{planck18}, with all values calculated for $H_0 = 70$~\kms~Mpc$^{-1}$.}.

\subsection{How is the cold interstellar medium distributed across the nearby galaxy population?}

The good agreement between the \hi\ and \hmol\ mass functions and a single Schechter function is in contrast with the stellar mass function in the nearby Universe, which is best characterised by a double Schechter function \citep{baldry12}. The double Schechter function, being the sum of two functions with the same $M^{\ast}$ but different $\alpha$ and $\phi_{\ast}$ values (as in Eq. \ref{eq:schechter}), better reproduces the total stellar mass function by separately accounting for the star-forming galaxies (abundant at low masses, with a high value for $\alpha$ and lower $\phi_{\ast}$) and the quiescent population (dominating at higher masses, with low $\alpha$ and higher $\phi_{\ast}$). 
If confirmed by more robust statistics at the low mass end, the fact that \hi\ and \hmol\ mass functions are well fitted by a single Schechter function suggests that, unlike stellar mass, most of the cold ISM mass in the nearby Universe is contained in a single galaxy population, namely star-forming galaxies as will be shown in Sec. \ref{sec:scalingrelations}. 

The values of \omegahi\ and \omegahmol\ imply that, overall, the molecular-to-atomic gas abundance ratio in the local Universe is 14.4\%. This is in stark contrast with the situation at $z\sim1.5$, the point in cosmic history when molecular gas was at its most abundant with \omegahmol$ / $\omegahi$\sim 75\%$  \citep{walter20}. 

There is also a stellar-mass dependence to the overall distribution of \hi\ and \hmol\ in the local Universe, with low mass galaxies (\mstar$<10^9$\msun) contributing between 30\% and 40\% 
of the overall \omegahi, but only 11\% of \omegahmol. Conversely, high mass galaxies (\mstar$>10^{10}$\msun) account for $\sim65\%$ of the molecular gas, and $\sim$40\% of all the \hi\ in the local Universe \citep{schiminovich10,lemonias13,hu20,fletcher21}. Such detailed measurements are proving very valuable to test and constrain cosmological hydrodynamic simulations, which while overall producing qualitatevely accurate \hi\ and \hmol\ mass functions and scaling relations, sometimes over- or under-predict the relative abundance of the two at either low or high stellar masses \citep[see][ for a detailed discussion of these matters for the SIMBA, Eagle and IllustrisTNG simulations]{dave20}.

\section{COLD GAS SCALING RELATIONS} 
\label{sec:scalingrelations}

We now turn our attention to the relations between global cold gas content and other galaxy properties, such as stellar mass, colour or star-formation rate -- these are generally referred to as {\it cold gas scaling relations}, and usually expressed in terms of gas mass (\mhi, \mh\ or \mgas) or gas-to-stellar mass ratio (\fhi $\equiv$ \mhi/\mstar, \fgas $\equiv$ \mh/\mstar\ or $f_{gas} \equiv$ \mgas/\mstar, often improperly called “gas fractions”). Because galaxy properties scale with mass, the latter formulation is usually preferred.

The usefulness of gas scaling relations was recognised early on, as a means of quantifying the typical \hi\ and \hmol\ gas content of galaxies and deviations from it \citep{robertshaynes94,young95}. In these studies, gas mass was typically normalised by luminosity in $B$-band or dynamical mass, 
and plotted as a function of luminosity, morphological type, colour, or other optical quantities. 

The study of gas scaling relations has undergone a renaissance in the past decade, thanks to the availability of large, representative samples of galaxies with \hi\ and \hmol\ measurements, along with a wealth of accompanying multi-wavelength data. Modern studies parameterise gas content in terms of stellar mass, stellar surface density, star-formation rate and other quantities that have an immediate physical interpretation and usefulness for theoretical models, facilitating the comparison with numerical simulations and semi-analytic models of galaxy formation and evolution. We review in Section \ref{sec:methods} the main approaches used to compute scaling relations in the presence of upper (or lower) limits and missing data.

\subsection{Methods}
\label{sec:methods}

\begin{figure}[h]
\includegraphics[width=3in]{ 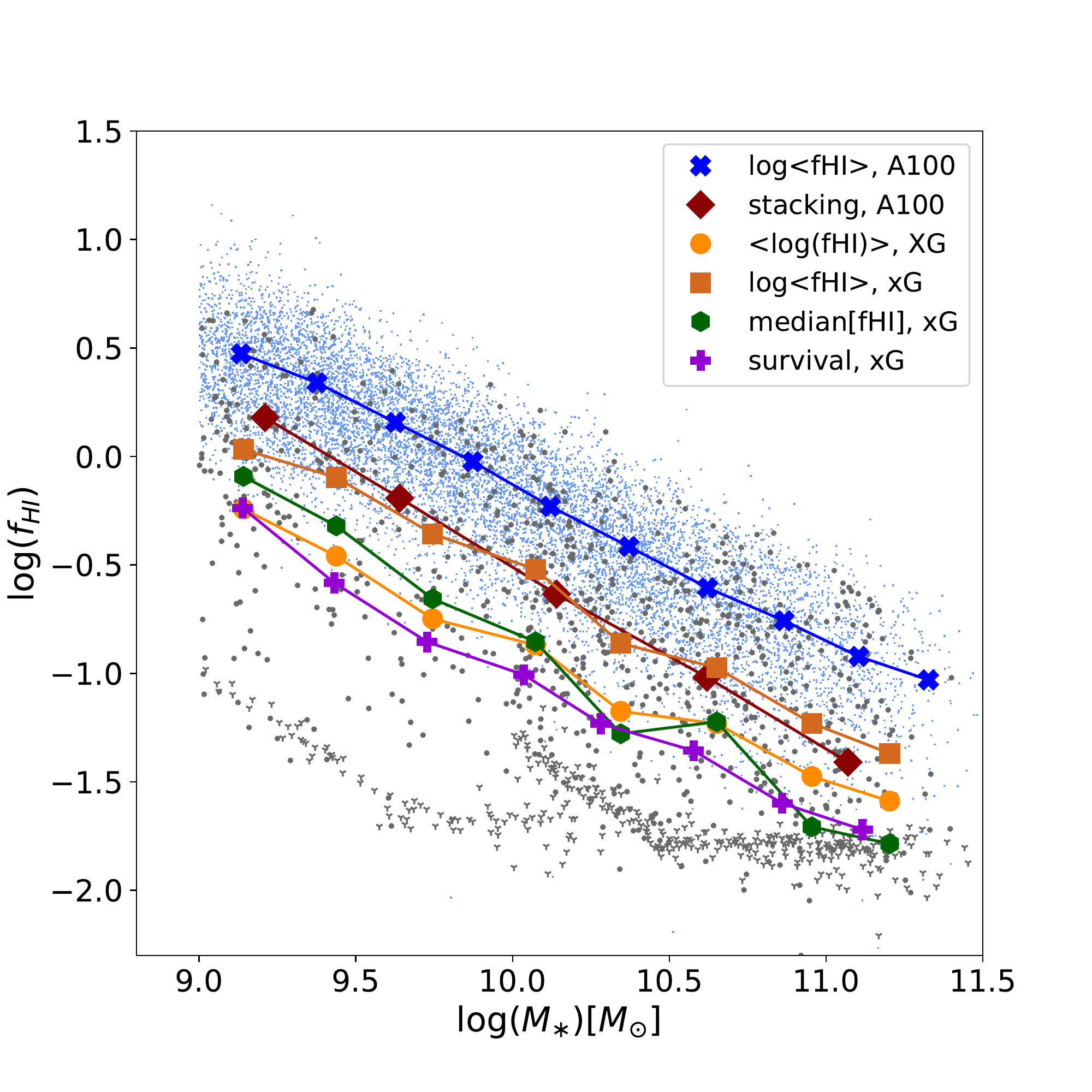}
\caption{Relation between \hi\ gas fraction and stellar mass for \hi\ detections (blue circles) from the ALFALFA survey (or ``A100", \citet{haynes18}, as referred to in the legend) and xGASS galaxies (gray, with downward arrows indicating non-detections). The scaling relation between \mstar-\fhi\ is derived in bins of \mstar\ using different methods: average \fhi\ of ALFALFA detections (blue), spectral stacking of ALFALFA data (red; mostly individual non-detections), for xGASS the difference between the logarithm of linear averages (orange) and average of logarithms (yellow), the median value of $\log$\fhi\ (green; including both xGASS detections and non-detections), and the logarithmic averages from survival analysis (purple).}
\label{fig:stacking}
\end{figure}

There are several aspects to keep in mind when it comes to quantifying gas scaling relations and comparing the results between different works. We discuss these below and refer to Figure~\ref{fig:stacking}, which shows \hi\ gas fraction as a function of stellar mass computed with different samples and/or methods to illustrate our main points. \\

\noindent
{\bf Sensitivity of observations}. Naturally, the sensitivity of the surveys used to compute scaling relations matters. For instance, unless one is interested in quantifying the average gas content of star-forming galaxies only, obtaining scaling relations from relatively shallow observations (such as wide-area blind \hi\ surveys) {\it using only detections} will return results that are biased toward high gas content, compared to samples that span the full range of galaxy properties, from star-forming to passive systems. This is shown by the blue line in Fig.~\ref{fig:stacking}, obtained by computing (logarithmic) averages of ALFALFA detections (blue dots).\\

\noindent
{\bf Averaging procedure}. There are several common estimators of the typical gas content, such as median and linear or logarithmic mean, which generally lead to different results, depending on the underlying distribution of the data. Hence, linear averages are usually offset from medians or logarithmic averages, reflecting the fact that galaxy properties typically follow log-normal distributions, and sometimes not even log-normal (see e.g. the bimodality of SFR in local galaxies). In Fig.~\ref{fig:stacking}, we use xGASS detections and upper limits (gray dots and arrows, respectively) to show the difference between logarithms of linear averages (orange line), logarithmic averages (yellow line) and medians (green line).\\

\noindent
{\bf Treatment of non-detections and upper limits}. Representative cold gas samples are typically selected from optical properties and followed up with radio and millimetre telescopes to measure the \hi\ and CO content, but not all galaxies are usually detected. Nonetheless, upper limits carry useful information that needs to be taken into account in the scaling relations, and there are three main approaches to do so. 

The simplest option is to compute {\bf medians}, rather than averages, to minimise the impact of the non-detections. Indeed, as long as the detection fractions are above 50\% and the upper limits do not significantly overlap with the detections (as is the case for the xGASS and xCOLD GASS surveys, which were designed to reach gas fraction limits of a few percent to produce stringent upper limits), medians are unaffected by the exact value of the gas mass of non-detections.

Another technique to deal with upper limits is {\bf spectral stacking}, which has become a common tool to constrain the statistical properties of a population of galaxies that lack individual detections in a survey. This turned out to be very powerful when applied to wide-area, blind \hi\ surveys, provided that independent spectroscopic redshifts were also available \citep[e.g.][]{fabello11a,brown15,brown17,meyer16}. Briefly, the \hi\ spectra of N galaxies are aligned in redshift and co-added (or {\it stacked}), regardless of whether these are individually detected or not, resulting in a total spectrum that yields an estimate of the total and average (when divided by N) \hi\ content of the galaxies that were co-added. The red line in Fig.~\ref{fig:stacking} shows the result obtained by stacking the ALFALFA \hi\ spectra from \citealt{brown15}. However, this technique has two important limitations. First, spectral stacking is an intrinsically linear operation, hence scaling relations from stacking cannot be blindly compared to ones  expressed in terms of averages of logarithmic quantities (e.g., compare the red and orange lines in the figure). This means that spectral stacking is most useful when used to compare {\it relative offsets} between subsets of galaxies  \citep[e.g., divided by specific SFR or halo mass;][]{brown15,brown17} rather than absolute values. Second, a stacked spectrum carries no information on the scatter of the underlying population whose spectra were co-added; while a handle on the variance of the result is usually obtained via the so-called {\it jackknife} statistics or by  binning the sample by multiple parameters simultaneously (if the size of the sample allows it), this remains challenging.

A third strategy to take non-detections into account is to adopt {\bf survival analysis} methods, which allow to compute linear regressions or means, medians and other statistical estimators for samples that include censored data such as upper limits, which are referred to as ``left-censored data" in survival analysis. We refer the reader to \citet{feigelson85} for an introduction of this branch of statistics to astronomers, and to \citet{feldmann19} for a discussion of existing approaches to compute likelihood functions in the presence of censored and missing data. These techniques have been applied to both \hi\ and \hmol\ datasets \citep[e.g.,][]{calette18,calette21,feldmann20}. Following \citet{feldmann19}, we show the averages for xGASS using survival analysis as a purple line in Fig.~\ref{fig:stacking}. The comparison with median values (green line) shows an offset of \about 0.2 dex between the two methods, especially at the low-mass end where upper limits are fewer. This is just a consequence of the fact that the underlying gas fraction distribution is not log-normal, hence medians and averages do not give the same answer.

Which technique is most appropriate depends on the sample used and on the specific question that one is trying to answer. Medians are meaningful only when upper limits are well separated from the detections and do not dominate the statistics in the bin. When upper limits and detections are mixed, survival analysis should be the preferred approach \citep[see e.g.][]{stark21}, considering that spectral stacking is intrinsically linear and thus should be used only to compare relative offsets rather absolute values. When bins are dominated by upper limits, determining scaling relations becomes more challenging, although survival analysis can recover means for detection fractions as small as 20\% (but with a large uncertainty; \citealt{calette18}). In this review, we compare medians to averages from survival analysis to demonstrate the impact of the non-detections on the observed trends.

\subsection{Gas fraction scaling relations} 
\label{sec:fgas}

The \hi\ and \hmol\ masses of galaxies correlate with several of their other global properties, such as luminosity, mass, size, morphology, and star formation rate \citep[e.g.][]{kenney89,sage93,robertshaynes94}. To illustrate our discussion of these results, Figures~\ref{fig:scaling_mass} and \ref{fig:scaling_sfr} show the main cold gas scaling relations from the xGASS and xCOLD GASS datasets as initially presented in \citet{xGASS} and \citet{saintonge17}.\footnote{While including other samples in these figures would extend the dynamic range in one or more dimensions (e.g., probing lower stellar masses, more gas-rich systems, interacting galaxies, early-type morphologies and so on), this would complicate the simple selection function of xGASS and xCOLD GASS and therefore the interpretation of any trends.
See however \citet{calette18} and \citet{ginolfi20} for more extensive data compilations.}

\subsubsection{Correlations with structural properties}

\begin{figure}[h]
\includegraphics[width=5in]{ 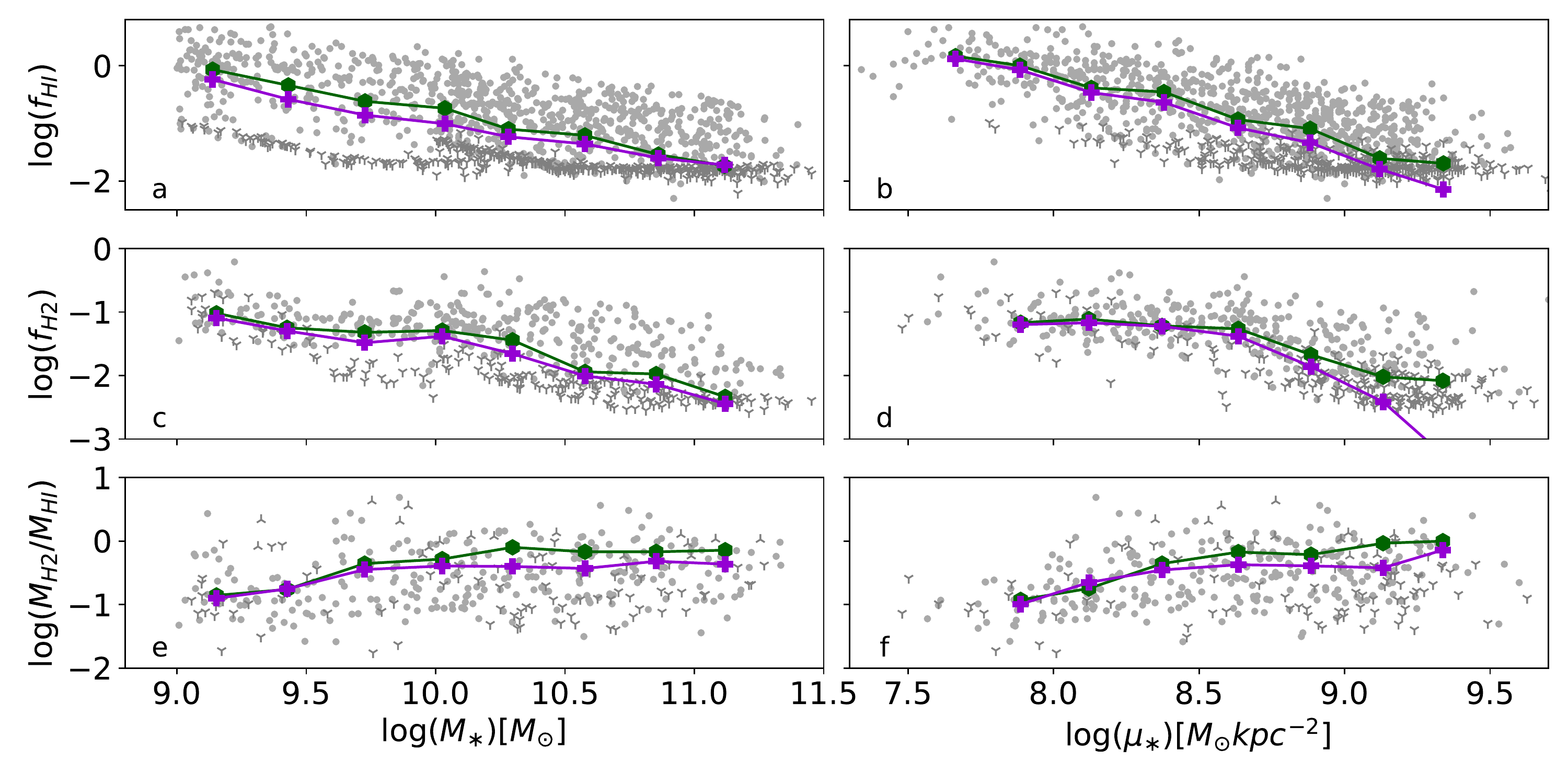}
\caption{Gas mass fraction scaling relations between the atomic gas mass fraction (\fhi$\equiv$\mhi$/{\rm M}_{\ast}$, top row), the molecular gas mass fraction (\fgas$\equiv$\mh$/{\rm M}_{\ast}$, middle row) and the molecular ratio (\mh/\mhi, bottom row) as a function of stellar mass (\mstar, first column), and stellar mass surface density (\must, second column). The gray dots and limits are the xGASS/xCOLD GASS galaxies, and the binned values are calculated as medians (green hexagons) and mean values using survival analysis (purple crosses) as described in Sec. \ref{sec:methods}.}
\label{fig:scaling_mass}
\end{figure}

Figure~\ref{fig:scaling_mass} shows the \hi\ gas fraction (\fhi$\equiv$\mhi$/{\rm M}_{\ast}$; top row), \hmol\ gas fraction (\fgas$\equiv$\mh$/{\rm M}_{\ast}$; middle row) and the molecular-to-atomic gas mass ratio (bottom row). These quantities are correlated with two basic structural properties: the stellar mass (first column) and the stellar mass surface density (\must$\equiv$\mstar (2$\pi r_{50,z}^2)^{-1}$). This quantity is commonly used as a quantitative measurement of the morphology of galaxies, with bulges growing in prominence as \must\ increases and becoming dominant at $\log$(\must/\msun kpc$^{-2}$)$>8.7$. 

Both \fhi\ and \fgas\ are anti-correlated with \mstar. Part of this trend is due to the increasing fraction of gas-poor, quiescent and mostly early-type galaxies as mass increases, although the trend is also found when considering only star-forming galaxies \citep{cicone17}. There is also an anti-correlation between \fhi\ and \fgas\ and \must, with galaxies becoming progressively more gas-poor the more bulge-dominated they are. Similar results are found when visual morphologies are used  \citep[e.g.,][]{boselli14_2}. 

Of note is the flattening of both the \mstar-\fgas\ (also noted by \citealt{jiang15}) and \must-\fgas\ relations (Fig. \ref{fig:scaling_mass}c\&d) below the characteristic thresholds of $\log$(\mstar)$\sim10.3$ and $\log$(\must)$\sim8.5$ where the galaxy population transitions from being dominated by quiescent, mostly early-type objects, to star-forming galaxies \citep{kauffmann03}. The lack of strong dependence of molecular gas contents on \mstar\ and \must\ in late-type systems (which dominate in Fig. \ref{fig:scaling_mass} at low \mstar\ and \must) has long been reported \citep[e.g.][]{young89a,casoli98,boselli14_2}. This break is however not seen in the equivalent \fhi\ scaling relations (Fig. \ref{fig:scaling_mass}a\&b). \citet{sage93} indeed showed that the \mh$/$\mhi\ ratio increases amongst spiral galaxies as the bulge component grows in importance (as in Fig. \ref{fig:scaling_mass}f). They however pointed out that this result was possibly of little physical significance since the excess \hi\ in low-mass, disc-dominated galaxies is located at large radii, away from the regions where most of the molecular gas is found -- we come back to this point in Section \ref{sec:applications}.

\subsubsection{Correlations with star formation activity}

\begin{figure}[h]
\includegraphics[width=5in]{ 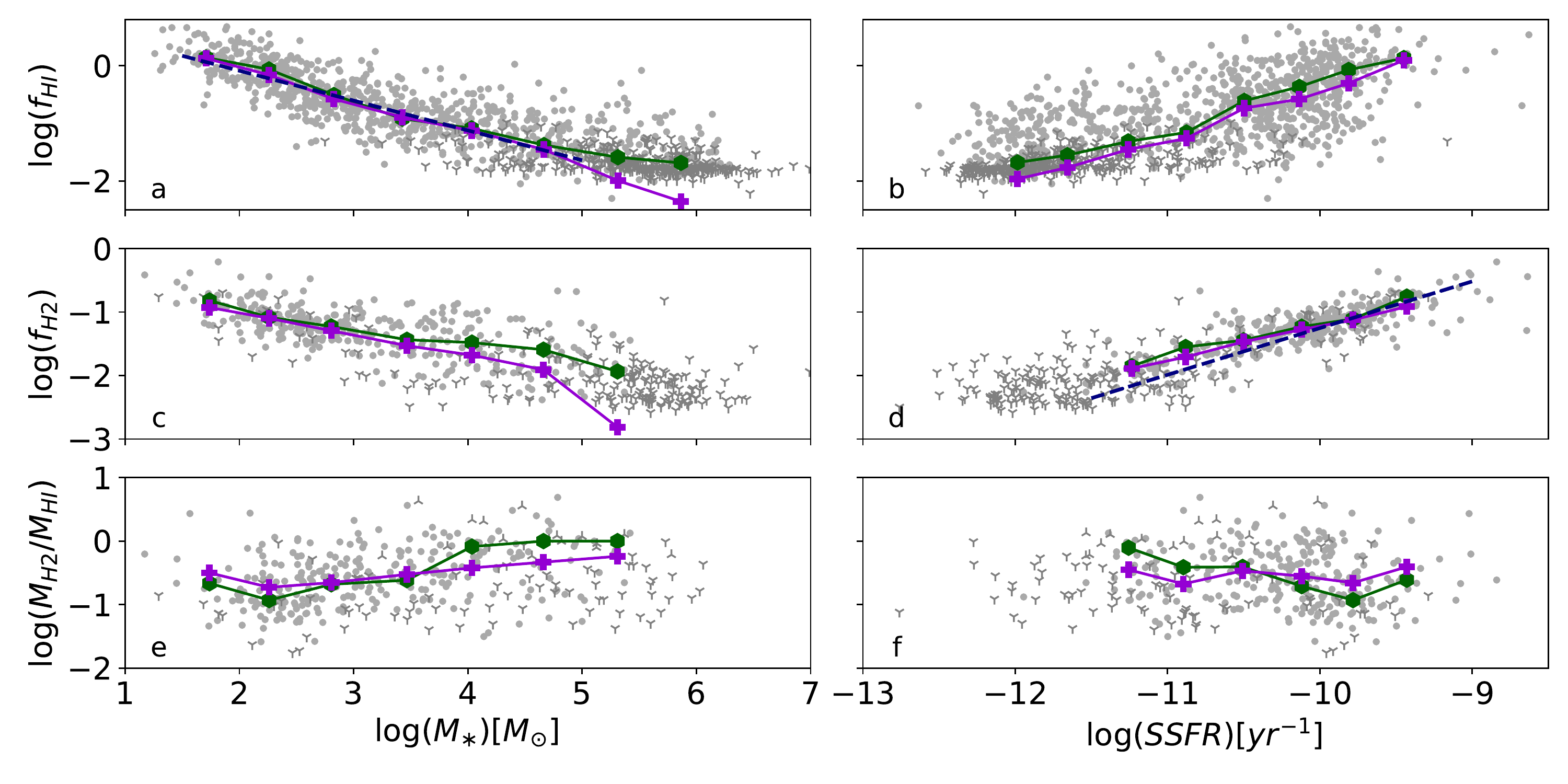}
\caption{Gas mass fraction scaling relations as in Fig. \ref{fig:scaling_mass} but for \nuvr\ color and specific star formation rate (SSFR). The dashed lines in panels a and d correspond to Equations \ref{eq:fhipred} and \ref{eq:fh2pred}, respectively.}
\label{fig:scaling_sfr}
\end{figure}

In Figure \ref{fig:scaling_sfr}, the gas quantities are shown as a function of two different estimates of the present-to-past averaged star formation activity, the NUV-r color and the specific star formation rate, SSFR. Both quantities compare a measurement of the current star formation activity (NUV flux in one case, the SFR in the other), with a measurement of the older stellar population (r-band flux and total stellar mass, respectively). The difference is in the modeling assumptions that have gone into the computation of SFR and \mstar, including a correction for dust extinction. 

The tightest of all the relations in Fig.~\ref{fig:scaling_sfr} is the one between \fgas\ and SSFR. This well-known correlation between molecular gas and star formation activity is often also portrayed as the relation between CO and total infrared luminosities \citep[e.g.][]{sanders85,solomon97} or as the relation between the surface densities of molecular gas mass and star formation rate \citep[i.e. the Kennicutt-Schmidt relation, e.g.][]{kennicutt98a,bigiel08}. In contrast, \fhi\ correlates more strongly with \nuvr\ than with SSFR. 
One might intuitively expect that \hi, being a long-term gas reservoir, would correlate more with a longer-term measure of SFR (i.e. traced by NUV) than a shorter-term one (e.g., traced by H$\alpha$ or FUV). However, the SSFRs in Fig.~\ref{fig:scaling_sfr} are based on \nuvr\ (and infrared) measurements, so this is not the case. Instead, this can be explained by the correlation between \hi\ and un-obscured star formation, which is evident in the outer parts of the star-forming disks, as illustrated convincingly by the case of M83 \citep{bigiel10}. More interesting than the correlation between gas and star formation itself however is the scatter of these relations, which reveals differences in global star formation efficiency across the galaxy population. This is discussed in Section \ref{sec:tdep}.

\subsection{Depletion time scaling relations}
\label{sec:tdep}

Once it has been established how much atomic and molecular gas is typically found in galaxies of a given type, we can investigate how the presence of that gas impacts on their star formation activity. This is most commonly studied through the global star formation efficiency, SFE, defined as the ratio between the SFR and the gas mass (either atomic, molecular, or total), or conversely the depletion time, $t_{dep} = ({\rm SFE})^{-1}$. The depletion time is an indication of the timescale over which star formation could be maintained at the current rate, given the available gas supply, and assuming no inflows of fresh fuel or recycling. In the nearby Universe, the typical depletion time for molecular gas in star-forming spiral galaxies is $\sim 1$ Gyr, and this value decreases slightly towards higher redshifts \citep{tacconi13,saintonge13}. These short depletion times ($<1$~Gyr) at look-back times of $>10$~Gyr are used as evidence for the steady accretion of gas onto galaxies in the intervening time, or at least for transport of atomic gas from the outer discs inwards; without such re-supply, there would be no gas and therefore no star formation in the Universe today, clearly at odds with observations. Another limitation of this simple interpretation of $t_{dep}$ is that not all the gas present can participate in the star formation process; this is particularly true of \hi, most of which does not have the density required for star formation and will remain as such in the absence of secular or external processes to change the status quo. 

\begin{figure}[h]
\includegraphics[width=5in]{ 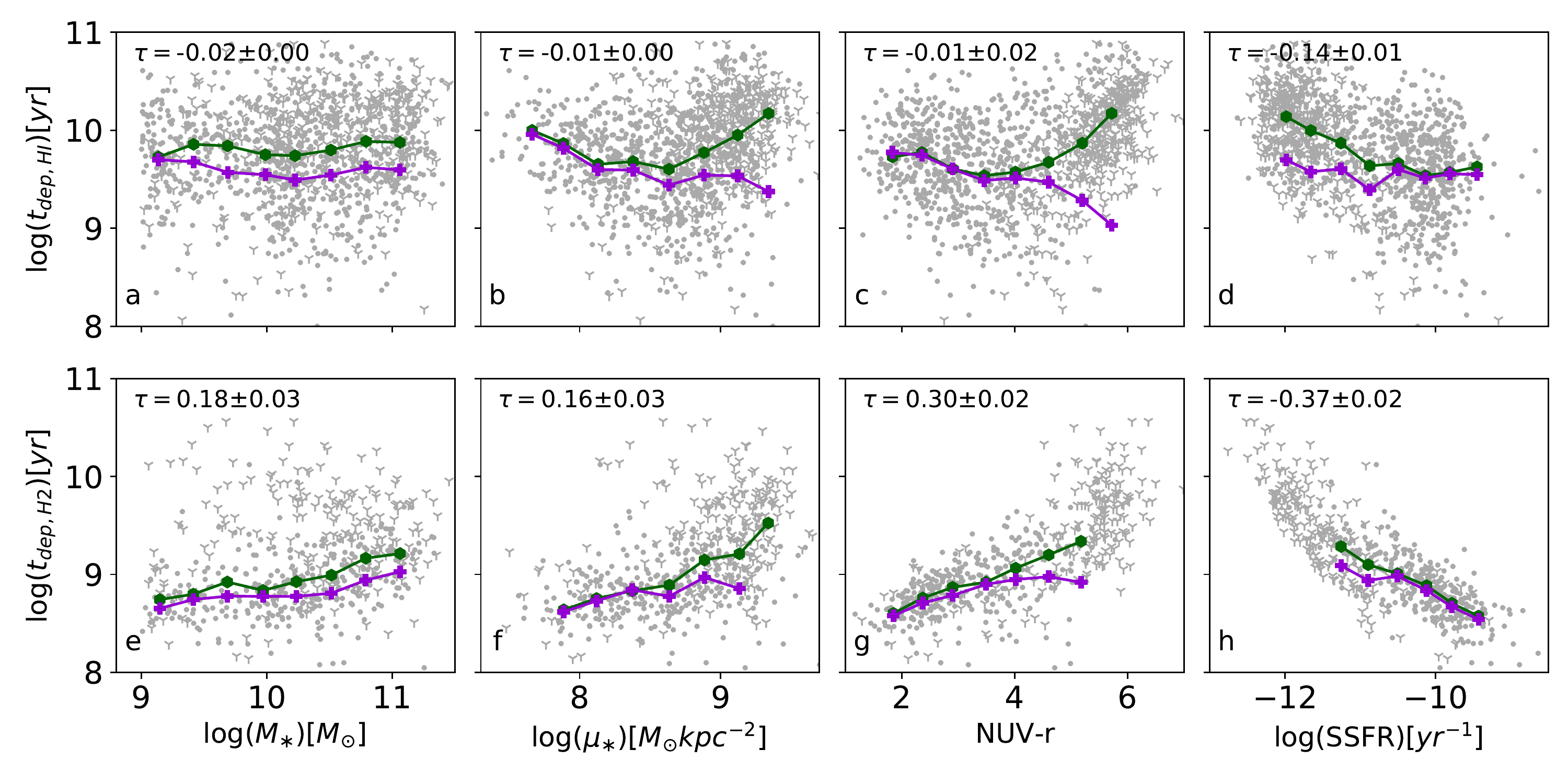}
\caption{Atomic and molecular gas mass depletion time scaling relations, with parameters, data, and symbols as defined in Fig. \ref{fig:scaling_mass}. In each panel, the Kendall rank correlation coefficient ($\tau$) is given.}
\label{fig:tdep}
\end{figure}

Figure \ref{fig:tdep} shows how the atomic and molecular gas depletion times vary as a function of the same four parameters used in Figures \ref{fig:scaling_mass} and \ref{fig:scaling_sfr}. Once upper limits are taken into account, \tdepHI\ shows no significant dependence on any of the quantities as judged by the Kendall $\tau$ coefficients, but rather is characterised by a large scatter of $0.5$~dex. The \hi\ distribution typically extends well beyond the optical disc, meaning that significant (but variable) amounts of the total \hi\ reservoir are not directly involved in the current star formation process. 

The molecular gas depletion time, \tdep, shows a different behaviour; it correlates with global galaxy properties (especially those with a direct link to star formation, like SSFR and \nuvr), and shows far less scatter. This is unsurprising given the much closer link between molecular gas and star formation, but it is a departure from the idea of a universal relation between molecular gas and star formation \citep[e.g.][]{bigiel11}, and a result that only emerged when parameter space was expanded sufficiently by surveys such as xCOLD GASS for these trends to appear. The results in Figure \ref{fig:tdep} suggest instead that a given amount of molecular gas does not always produce the same levels of SFR; rather, the gas-to-star conversion efficiency appears to vary systematically across the galaxy population. Once combined with the variations of \fhi\ and \fgas, the implication of Figure \ref{fig:tdep} is that the specific level of star formation activity in galaxies is determined by both the amount of gas available, and the efficiency of the conversion of that gas into stars \citep[e.g.][]{saintonge16}. 

The dependence of \tdep\ on global galaxy properties is at first puzzling; how can the details of star formation, a process occurring on parsec scales, be ``controlled" by the global properties of galaxies on scales of tens of kpc? Or could this apparent correlation be caused by another common factor? These questions are best answered by looking at the Kennicutt-Schmidt (KS) relation. On global scales, the \tdep\ variations shown in Fig. \ref{fig:tdep} manifest themselves as systematic deviations of particular galaxies. For example, bulge-dominated galaxies (e.g. those with high values of \must\ in Fig. \ref{fig:tdep}) tend to fall systematically below the KS relation traced by spiral galaxies \citep{COLDGASS2,davis14}. At $\sim$~kpc resolution, the variations in \tdep\ are found to be larger from galaxy to galaxy than within individual galaxies; in other words, individual galaxies follow their own KS relation, but the exact slope and normalisation of this relation varies systematically based on global galaxy properties \citep{ellison21}. Variations are even seen at the Giant Molecular Cloud (GMC) level. In nearby galaxies where molecular gas has been mapped on scales of tens of pc, GMCs are found to be systematically bigger/brighter in some galaxies \citep{hughes13}. What's more, these galaxy-to-galaxy variations in cloud properties are linked to global galaxy properties, such as mass and morphology, via changes in internal pressure and structure \citep{sun20,rosolowsky21} and therefore systematic variations in the amount of dense star-forming gas compared to the total (diffuse) molecular gas mass traced by CO(1-0). This is discussed further in Section \ref{sec:tdepvariations}.

\subsection{Application of scaling relations: estimating cold gas masses} 
\label{sec:predictors}

Gas fraction scaling relations, like those shown in Figures \ref{fig:scaling_mass} and \ref{fig:scaling_sfr}, give us expectation values for the gas contents of ``normal" galaxies, which is of crucial importance to study the impact of additional factors such as presence of an AGN, interactions, or environment. For example, the tight correlation between \mhi\ and optical isophotal diameter at fixed morphological type \citep{hg84} was of particular importance early on as it led to the definition of the \hi\ deficiency parameter to quantify environmental effects and show how clusters impact on the ISM of infalling galaxies \citep{giovanelli85,dawes9}. 

Stellar mass by itself is a relatively poor predictor of gas content, with a spread of over 2 orders of magnitude in \fgas\ and \fhi\ at fixed \mstar\ (Fig. \ref{fig:scaling_mass}a\&c). This is especially true for high mass galaxies (\mstar$>10^{10}$\msun), and unless additional information can be used to first break the degeneracy between star-forming and quiescent galaxies at fixed mass \citep{brown15}. There are however a number of strategies at our disposal:  

\subsubsection{Control samples} The first approach is to use a large representative galaxy sample with direct \hi, CO (or dust) measurements, and to draw control samples from it. The \hi\ and \hmol\ masses of a galaxy of interest can be estimated to be the mean of the gas masses of the galaxies most similar to it in the reference sample. The choice of which parameters to control against depends on the specific science goals and the properties of the sample, but the general gas scaling relations can guide those choices. Given all the correlations presented in  Figures \ref{fig:scaling_mass} and \ref{fig:scaling_sfr}, it is recommended to select control galaxies by matching with the target object on two parameters, one that controls for mass (either \mstar, or luminosity in a NIR band) or alternatively morphology (\must\ for example), as well as a parameter that traces star formation activity (SFR itself, or luminosity in the UV or FIR). The combination of \mstar\ and SFR is most often used to select control samples to establish how specific phenomena impact the gas contents of galaxies \citep[e.g.][]{violino18,ellison19,koss21}. 

\subsubsection{Empirical relations} 
Alternatively, empirical scaling relations can be used to produce expectation values of \mhi\ and \mh. These methods are best used in the same spirit as the control sample technique described above. If empirical estimates of \mhi\ and \mh\ are nonetheless used in lieu of direct measurements (see Section \ref{sec:indirecttracers}), caution should be taken (1) to take into account the intrinsic scatter of the relation used, (2) to stay within the parameter space where the relations have been calibrated, and (3) to be mindful of underlying assumptions in the calibration, which may limit the applications. 

\paragraph{Inferring \mh\ from the SFR}

Of all the scaling relations shown in Figures \ref{fig:scaling_mass}$-$\ref{fig:tdep}, those directly relating \mh\ and SFR are the tightest. This strong correlation has long been exploited to infer \mh\ values, a technique sometimes referred to as ``inverting the Kennicutt-Schmidt relation". Based on the xCOLD GASS data (Fig. \ref{fig:scaling_sfr}d) the recommended relation to predict \mh\ values for galaxies in the local Universe is: 
\begin{equation}
    \log ( M_{H_2} / M_{\ast}) = (0.75\pm0.01) \log({\rm SSFR}) + (6.24\pm0.12). 
    \label{eq:fh2pred}
\end{equation} 
A factor of $\log_{10}(1.36)$ should be added to the right hand side to obtain the total molecular mass fraction, $\log(M_{mol}/$\mstar$)$, including the contribution of He and metals. This relation is best applied for galaxies with $\log({\rm SSFR})>-11.5$ where it is robustly constrained by observations. For CO detections, the scatter around this relation is 0.24 dex, of which 0.12 dex is intrinsic, and 0.20 comes from measurement uncertainties. By combining local and high redshift observations,  \citet{tacconi20} propose a calibration of this relation that is applicable at all redshifts up to $z\sim3$. Others have fitted the 3D parameter space formed by \mh, \mstar\ and SFR to provide redshift-independent gas predictors \citep{santini14}. 

For local Universe galaxies, the good availability of mid-infrared photometry from the WISE all-sky survey \citep{wright10} at wavelengths of 3.4, 4.6, 12, and 22 $\mu$m can also be exploited to estimate \mh\ values. The two shorter wavelength bands trace the established stellar population, while the 12, and 22 $\mu$m bands probe warm dust and Polycyclic Aromatic Hydrocarbon (PAH) emission, therefore being sensitive to the ongoing rate of star formation. \citet{yesuf17} have indeed shown that the flux ratio between 12 and 4.6$\mu$m correlates well with \fgas.  This is explained by the fact that the 12-to-4.6$\mu$m flux ratio is a proxy for SSFR, which itself has been shown to correlate tightly with \fgas\ (see Section  \ref{sec:scalingrelations}).  A tight correlation ($r_s\sim0.9$) is also found directly between \mh\ and $L_{12\mu m}$ \citep{jiang15}. This is expected for star-forming galaxies where $\sim80\%$ of the 12$\mu$m emission is due to stellar populations younger than 600 Myr \citep{donoso12}, but caution must be applied in quiescent galaxies where older stellar populations contribute significantly, breaking the relation between MIR emission and molecular gas \citep{davis14}. \citet{gao19} expand on this by showing that adding \mstar\ and $g-r$ to $L_{12\mu m}$, yields predicted $L_{CO(1-0)}$ values with 0.16 dex of intrinsic scatter compared to direct measurements, an improvement from the 0.21 dex of scatter obtained by using $L_{12\mu m}$ only. 

When estimating \mh\ values from the SFR, the implicit assumption of a constant molecular gas depletion timescale is made. While this is a reasonable assumption for ``normal" star-forming galaxies, we showed in Section \ref{sec:tdep} that it is not in other regimes. The scope of analysis possible with \mh\ values predicted from SFRs is also limited, because of these in-built assumptions. 

\paragraph{Inferring \mhi\ from scaling relations}
Scaling relations have also long been used to estimate \mhi, with the community initially focussed on environmental studies at $z=0$ via the \hi\ deficiency parameter, using offsets from the scaling relations to identify systems affected by ram pressure or other environmental processes. As mentioned above, \hi\ deficiency was originally defined in terms of optical diameters, but over time other parameterizations based on colour and structural properties, broad-band luminosity, stellar mass or specific angular momentum have been used to various degrees of success (see  \citealt{dawes9} for a critical comparison and a discussion of the limitations of these different approaches). More recently, ``photometric gas fractions'' estimates have been used to increase the size of $z=0$ samples by taking advantage of large multiwavelength surveys in the UV and IR regimes \citep[e.g.,][]{kannappan04,zhang09,eckert15}.

In the context of this review, and to compare with the above \hmol\ results, Fig. \ref{fig:scaling_sfr}a shows that \nuvr\ color is the best single-parameter predictor of \fhi, with a scaling of: 
\begin{equation}
    \log ( M_{HI} / M_{\ast}) = (-0.52\pm0.05) ({\rm NUV}-r) + (0.95\pm0.02).   
    \label{eq:fhipred}
\end{equation} 
The scatter of \hi\ detections around this best-fit relation is however significant, at 0.39 dex. We also caution that this relation is best used when \nuvr$<5$, the regime where observations provide the best constraints. 
Other \hi\ gas fraction predictors in the literature adopt linear combinations of properties such as a colour (\nuvr\ or $g-r$) and \must\ to correct for a second-order dependence on morphology that contributes to this dispersion \citep[e.g.,][]{zhang09,GASS1}, finding scatters of order of \about 0.3 dex. In an interesting work, \citet{teimoorinia17} explore the application of (non-linear) artificial neural networks to estimate \hi\ gas mass fractions from SDSS properties, based on a training set of galaxies with ALFALFA detections or upper limits. This technique improves on linear methods by yielding a lower scatter of 0.22 dex, but does not reproduce the gas fractions of data sets with significantly different selection such as GASS \citep{GASS8} without additional cuts, demonstrating once again that ``what you get out is what you put in''.

\section{THE COLD ISM AND GALAXY EVOLUTION}
\label{sec:applications}
The general purpose of galaxy evolution studies is to understand what regulates star formation, allowing galaxies to grow differently over time, resulting in the full range of galaxy properties observed  (e.g. masses, sizes, morphologies, metallicities and kinematics). Particularly useful are the scaling relations between these properties, as they provide strong constraints on the mechanisms driving galaxy evolution. For example, it is not enough to be able to explain how we can form both spiral and elliptical galaxies, the same physics must simultaneously explain why spiral galaxies tend to be star-forming while most elliptical galaxies are not. The properties and scaling relations used to constrain galaxy evolution processes compare quantities such as stellar masses, colours, morphologies, metallicities, or environment. Our purpose here is to illustrate how adding information about the cold ISM clarifies the picture. 

\begin{figure}[h]
\includegraphics[width=5in]{ 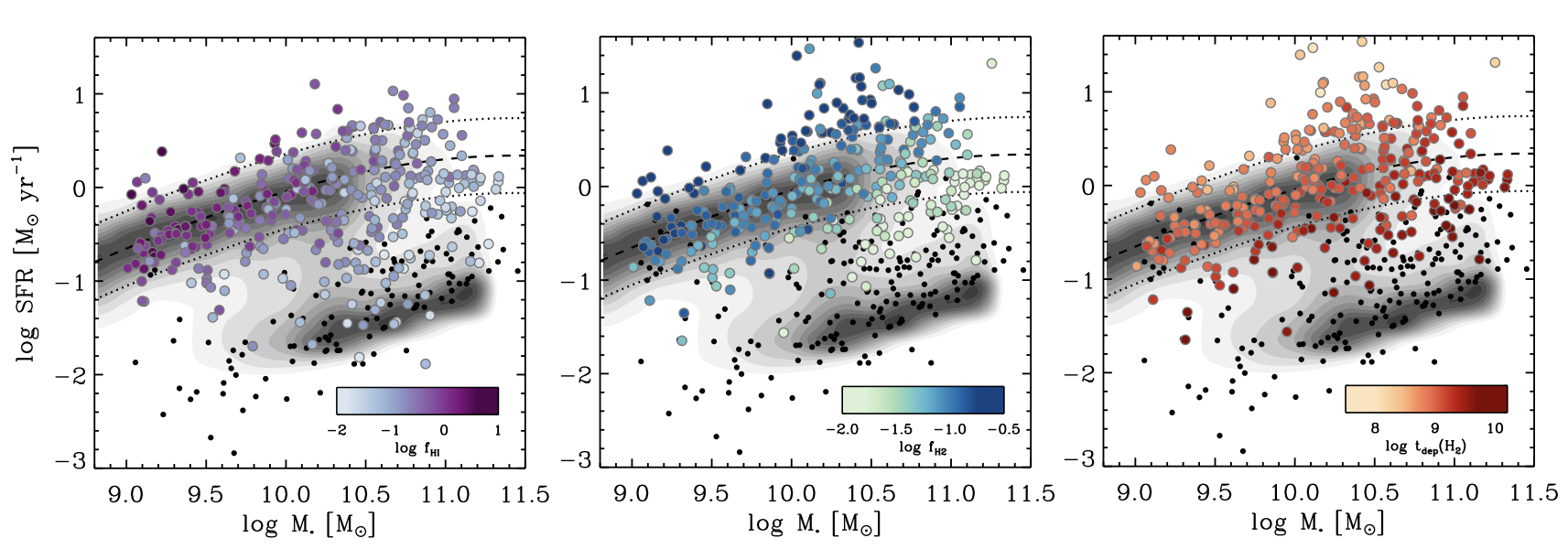}
\caption{Variations in \fhi, \fgas\ and \tdep\ across the \ms\ plane. Small black dots are non-detections (i.e. gas fractions are below the xCOLD GASS survey limit of $\sim2\%$), and the contours show the full SDSS sample at $0.02<z<0.05$. Note that the apparent hard lower limit of the passive cloud is an artefect due to the sensitivity of the observations and the SFR calculations, with the true SFRs of the quiescent object in fact a continuum to zero.  The dashed and dotted lines indicate the MS and the $\pm$0.4 dex scatter around it, respectively. Figure reproduced from  \citet{saintonge17}.}
\label{fig:MS_gas}
\end{figure}

We focus our discussion on the distribution of galaxies in the 2D parameter space formed by stellar mass and star formation rate. This picture has replaced the color-magnitude diagram as the de facto tool to represent the evolutionary state of galaxies; it highlights the bimodality of the galaxy population, and allows samples of galaxies to be put in their larger context. Figure \ref{fig:MS_gas} shows the distribution of the xCOLD GASS sample in the SFR-\mstar\ plane, with the different panels showing how the atomic and molecular-gas mass fractions as well as \tdep\ vary. These are ``2D" versions of the scaling relations presented in Sec. \ref{sec:scalingrelations}, so unsurprisingly similar observations are readily made: amongst the star-forming galaxies, \fhi\ varies most with \mstar\ while \fgas\ and \tdep\ have a stronger dependence on SSFR. 

\begin{marginnote}[]
\entry{MS}{Main Sequence, defined as the locus of star-forming galaxies in the 2D parameter space formed by \mstar\ and SFR.}
\entry{\deltams}{The offset of a galaxy from the main sequence, defined as $\log{\rm SFR}-\log{\rm SFR_{MS}}(M_{\ast})$.}
\entry{SFH}{Star Formation History, the time-dependent rate of star formation of a galaxy over its lifetime.}
\entry{SSFR}{Specific Star Formation Rate, the rate of star formation per unit stellar mass, i.e. SFR/\mstar. }
\end{marginnote}

\subsection{The star formation main sequence}

The dominant feature in the SFR-\mstar\ plane is the galaxy main sequence (MS), the tight relation between \mstar\ and SFR traced by star-forming galaxies. The general observational consensus is that the MS is $\sim$linear up to a critical mass of  $M_{knee}$, and flattens out towards higher masses. The MS has been observed up to $z\sim4$, with the MS shifting towards higher SFRs as $\sim(1+z)^2$ \citep[e.g.][]{lilly13,whitaker14,tomczak16}, while the scatter around the MS of $\sim0.2-0.3$~dex depends only weakly on \mstar\ and $z$ \citep{speagle14,popesso19}.  

If the MS were perfectly linear and scatter-free at all masses, we would conclude that galaxies are self-similar, with more massive systems having more gas, and therefore being able to sustain higher levels of star formation activity. However, explaining the observed shape (i.e. the low-mass slope and the flattening above $M_{knee}$) and the amount of intrinsic scatter requires careful consideration of what sets the amount of gas available for star formation under different conditions, and the efficiency of the conversion of this gas into stars\footnote{The third key feature of the observed MS, its redshift evolution and the link with the gas contents of star-forming galaxies across cosmic time, is reviewed in detail in \citet{tacconi20}.}. 

The exact functional form and normalisation of the $z\sim0$ MS is sensitive to methodology and selection effects, as well as the calibrations used to calculate the \mstar\ and SFR values \citep[e.g.][]{popesso19}. We use the MS definition of \citet{saintonge16}, but slightly revisited by using the \mstar\ and SFR values from the GSWLC-2 \citep{salim18}, and the fitting function of \citet{lee15}. The best fit function we adopt, based on SDSS galaxies with $0.01<z<0.05$ is: 
\begin{equation}
    \log {\rm SFR_{MS}}~[M_{\odot}~{\rm yr}^{-1}] = 0.412 - \log \left( 1 + \left[ \frac{10^{\log(M_{\ast})}}{10^{10.59}} \right]^{-0.718} \right)
    \label{eq:msfit}
\end{equation}

\subsubsection{The shape of the main sequence}

A common explanation for the flattening of the $z\sim0$ MS above $M_{knee}\sim3\times 10^{10}$\msun\ is the increased importance of stellar bulges  \citep{erfanianfar16,frasermckelvie21}. The same phenomenon also naturally explains why the flattening decreases in strength as redshift increases \citep{lee15,leslie20}. In the local Universe and on the MS, while galaxies are always rotationally-dominated systems \citep{frasermckelvie21}, the frequency of stellar bulges indeed increases with \mstar, as shown by the higher stellar mass surface densities \citep{saintonge16} and higher bulge-to-total mass ratios \citep[from $B/T\sim0$ at \mstar$=10^9$\msun\ to 0.3 at \mstar$=10^{11.5}$\msun;][]{cook20}. The incidence of strong bars also increases along the main sequence \citep{gavazzi15}. 

A simple explanation for these observations would be that while a bulge contributes to the stellar mass, it does not (typically) contribute to the SFR, explaining the flattening of the MS as bulges become more dominant. However, \citet{cook20} find that when building a main sequence only accounting for stellar mass in the disc component of galaxies, the overall shape of the main sequence does not significantly change. Another factor must be responsible for making even the discs of massive main sequence galaxies less star-forming.

\begin{figure}[h]
\begin{subfigure}[h]{0.5\linewidth}
\includegraphics[width=\linewidth]{ 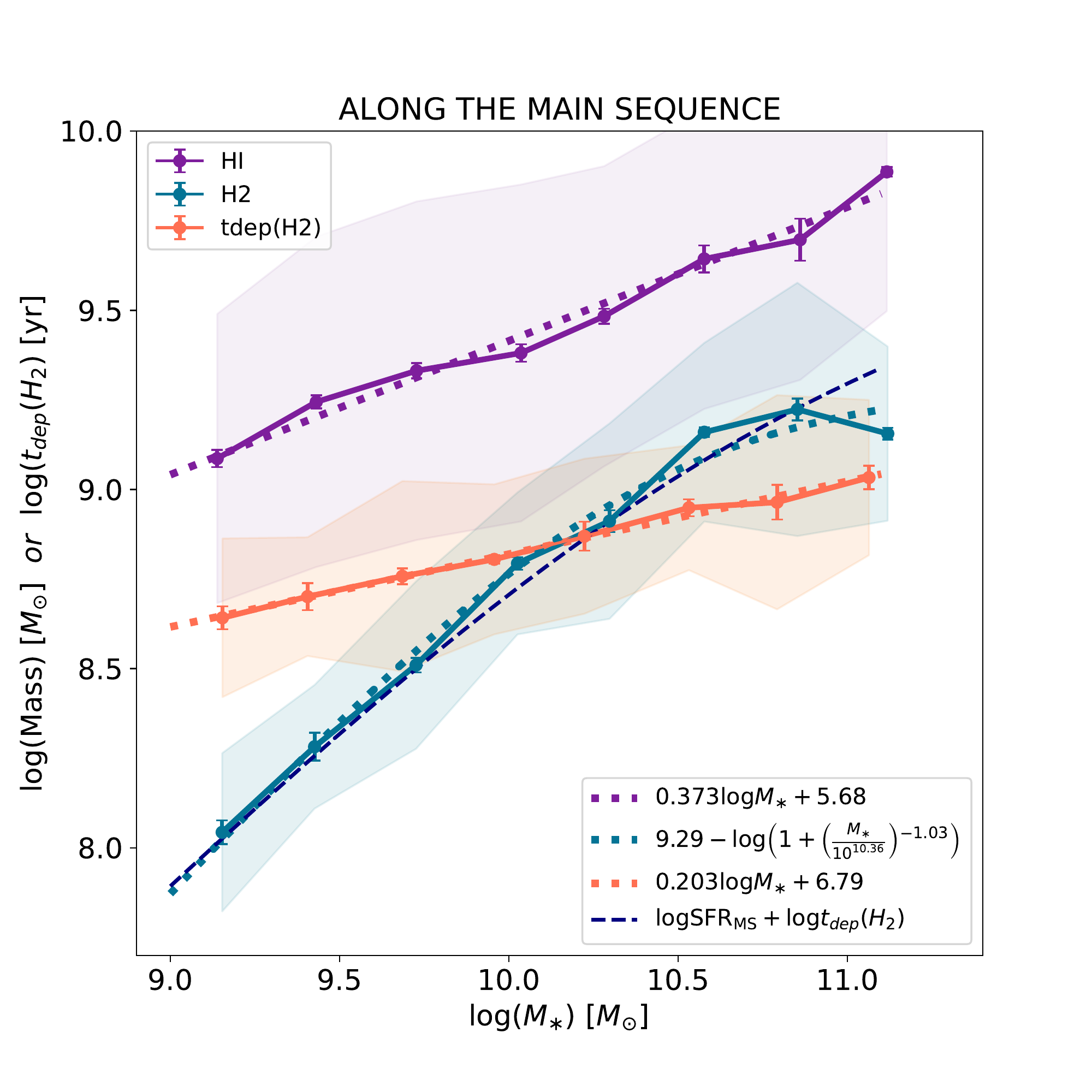}
\end{subfigure}
\hfill
\begin{subfigure}[h]{0.5\linewidth}
\includegraphics[width=\linewidth]{ 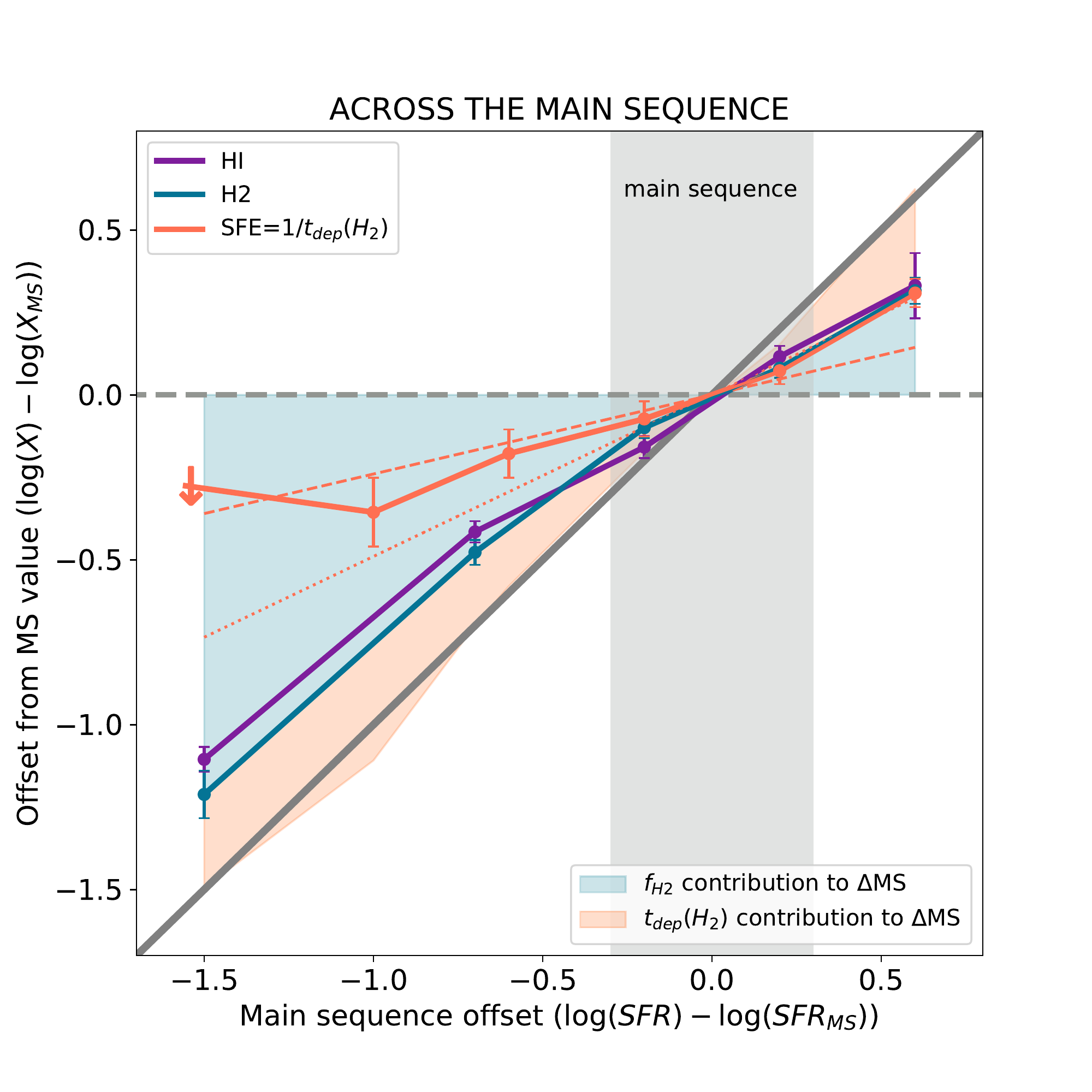}
\end{subfigure}%
\caption{{\bf  Left:} \mh, \mhi\ and \tdep\ as a function of stellar mass for main sequence galaxies. Error bars are 1$\sigma$ confidence intervals on the position of the mean values from bootstrapping, while shaded areas show the $1\sigma$ scatter. Best fitting relations are shown as dashed lines and their equations given in the caption. The dashed navy line shows the predicted log(\mh) obtained by combining Eq. \ref{eq:msfit} and the mass-dependence of \tdep\ on the main sequence (i.e. the dashed orange line). {\bf Right:} Mean global star-formation efficiency (SFR/\hmol), \mh\ and \mhi\ as a function of the offset from the main sequence, normalised to the value of each quantity on the MS. Also shown are the results for the variation of \tdep\ with MS offset from \citet{tacconi20} (dotted line) and \citet{feldmann20} (dashed line). The diagonal gray line shows the increase of the SFR; if either \mh, \mhi\ or \tdep\ alone were responsible for explaining \deltams, the corresponding data would fall on this 1:1 line.}
\label{fig:alongacross}
\end{figure}

We can also seek a gas-centric explanation for the shape of the MS. The flattening would be naturally explained if, as galaxies grow more massive along the MS, their cold gas reservoirs do not increase proportionally (or alternatively, their star formation efficiency decreases), causing a reduction of the SFR per unit stellar mass. The argument has indeed been made that the MS is a consequence of the more fundamental relations between \mh\ and \mstar\, and the tight relation between \mh\ and SFR \citep{lin19,feldmann20}. This is also illustrated in Figure \ref{fig:alongacross}a. For main sequence galaxies, log\mh\ rises $\propto 0.79\log$\mstar\ up to $M_{knee}\sim10^{10.3}$\msun\ and flattens above it (blue line and points), while \tdep\ increases monotonically as $\propto 0.2\log$\mstar\ (orange line and points). Since any molecular gas present in a bulge-dominated system forms stars at a lower efficiency \citep{martig09,davis14}, the increased fraction of bulge-dominated galaxies on the MS with mass may explain this global increase of \tdep.  This may therefore be the main contribution bulges make to the shape of the MS. However, we note that the dependence of \tdep\ on \mstar\ continues below 10$^{10}$\msun, suggesting that stellar bulges are most likely not be the only/main player in regulating the ability of cold gas to feed star formation.

Combining the slopes from these fits produces a shape and slope for the main sequence that is in agreement with other studies at low redshifts  \citep[see e.g. the compilation of ][]{speagle14}. For illustrative purposes, the reverse is shown in Figure \ref{fig:alongacross}: the scaling of \mh\ for main sequence galaxies can be accurately predicted from $\log{\rm SFR_{MS}}$ and \tdep\ (navy dashed line). 

Having been able to ascribe most of the shape of the MS to the molecular gas contents of galaxies, our attention should turn to the mechanisms that regulate this gas availability. First, we observe that \mhi\ increases along the main sequence as $0.37\log$\mstar\ (Figure \ref{fig:alongacross}a, purple line). The shallowness of this relation and the significant scatter ($\sigma=0.47$) once again highlight the looser link between \hi\ and star formation. It also suggests that the mechanisms responsible for fuelling galaxies with gas and setting what fraction of the total cold gas reservoirs satisfy the requirements for molecule and star formation are fundamentally responsible for setting the shape of the MS. 

For a standard stellar-to-halo mass ratio, the mass at the knee of the MS, $\sim3\times 10^{10}$\msun\ corresponds to a halo mass of $\sim10^{12}$\msun. Tantalisingly, this is the threshold mass found in simulations, where accretion of gas transitions from cold to hot mode \citep[e.g.][]{keres09}. The cold mode is key to sustain the molecular gas reservoir and star formation, as the gas accreted through this channel is cooler, denser, and most of its mass ends up in the central star-forming disc \citep{vandevoort12}. This picture is also consistent with the observation that even the disc component of MS galaxies above the knee is more gas-poor and less star-forming \citep{popesso19,cook20}. 

An alternative explanation is that the internal mechanics of galaxies (also referred to as ``secular processes") regulate the gas availability. For example, \citet{gavazzi15} suggest that bars may play a crucial role in reducing star formation activity in the discs of massive MS galaxies. In their scenario, star-forming galaxies above the knee of the MS tend to experience bar instability, which funnels gas to the central region in a few dynamical times, triggers a nuclear starburst, the formation of a (pseudo) bulge, and subsequently a drop in the star formation activity. This scenario is consistent both with the higher fraction of bulges on the MS at high masses, and the lower gas fractions.

\subsubsection{The scatter of the main sequence}

The scatter around the main sequence is a consequence of the star formation histories of individual galaxies at a given stellar mass. Understanding the nature, evolution (both with mass and redshift), and the timescales associated with this scatter is a currently active area of research \citep[e.g.][]{speagle14,matthee19,berti21,sherman21}. 

There are (at least) two competing ideas to explain the scatter about the star-formation main sequence. In the first, galaxies have SFHs that vary on short timescales, resulting in galaxies constantly moving up and down with respect to the MS. In the second, galaxies have less ``eventful" SFHs, resulting in their offset from the MS to persist on long timescales.  The first model is supported by the ``equilibrium" model for galaxy evolution \citep[e.g.][]{lilly13} where galaxies regulate their star formation activity via gas inflows and outflows. In this model, a gas accretion event will be followed by an increase in star formation activity. The intense star formation will drive an outflow, depleting the gas reservoir further and forcing the SFR back down. The scatter of the MS is therefore linked to the time required for galaxies to return to their equilibrium state after a perturbation of their gas contents \citep{finlator08}. Based on simulations, \citet{tacchella16} estimate that galaxies would oscillate around the MS on timescales of $\sim0.4 t_{\rm Hubble}$ (i.e. $\sim5$~Gyr at $z=0$, but as little as $\sim1$~Gyr at $z\sim3$).

Other simulations however suggest that galaxies ``remember" their star formation histories, with the present-day SFRs connected to the formation time and halo properties \citep{matthee19}. Under such a scenario, the offset of a galaxy from the MS varies on long timescales ($\sim10$~Gyr), and is related to its position in the cosmic web. Observational evidence for this in the nearby Universe has been found by \citet{berti21}, who report that at fixed \mstar, galaxies above the MS are less clustered that those below. Using MaNGA \citep{bundy15} data and reconstructing the star formation histories, Bertemes et al. (2021, in prep) find a similar result: the majority of galaxies at $z\sim0$ are found on the same ``side" of the MS where they have been on timescales of Gyrs, with short-term stochasticity only accounting for a small fraction of the scatter around the MS. 

The observation of a systematic difference in \fhi\ and \fgas\ between galaxies in the upper and lower half of the MS scatter (see Figures \ref{fig:MS_gas} and \ref{fig:alongacross}b) suggests that the scatter of the main sequence is gas-driven. It also argues against the rapid motion of galaxies around the MS, rather suggesting that MS offsets persist on timescales of \about~Gyrs. The result for \hi\ is particularly constraining; with such a long depletion time (\tdepHI$\sim5$~Gyr, see Fig. \ref{fig:tdep}), the \hi\ reservoir is sluggish and does not respond quickly. Rapid variations in the SFHs on $\sim$Gyr timescales are therefore not expected to be seen in the global \fhi\ and \fgas\ values, and would rather work towards erasing the scaling between gas mass fraction and SSFR.  Comparing all the various timescales that influence the baryon cycle and the star formation process, \citet{tacconi20} argue that since cosmic noon ($z<2$) the star formation activity of galaxies is limited by the rate of gas accretion rather than by star formation efficiency. This is in line with the scenario where the offset from the main sequence is linked to halo properties, as those regulate gas availability and therefore star formation histories.

\subsection{Above the main sequence}
\label{sec:aboveMS}

Galaxies in the local Universe can have star formation rates as high as hundreds of solar masses per year, putting them two orders of magnitude above the main sequence. Major mergers are the process most likely to significantly push galaxies well above the main sequence, and indeed the highest SFRs in the local Universe are found in merging gas-rich galaxies at the time of coalescence of their nuclei \citep{sanders96, larson16}. 

Major mergers at $z\sim0$ are very rare events, making them minor contributors to the overall ``star formation budget", despite the high SFRs they can generate. Far more significant to this budget are the processes that may elevate galaxies above the main sequence by as little as a factor of $\sim 2-3$, but occur far more frequently, such as minor mergers, galaxy interactions, and bar instabilities.  

Taking a gas-centric view, there are two possible channels to explain why dynamical processes such as these can elevate the SFR of galaxies: (1) they may increase the amount of gas available to participate in the star formation process, and (2) they may change the properties of the gas, resulting in an increased star formation efficiency (i.e. more stars formed per unit total gas mass). Figure \ref{fig:alongacross}b makes the case for both of these factors to be at play, with starbursting galaxies (defined here as those more than 0.4 dex above the MS) having excess \hi\ and \hmol\ and shorter \tdep.

\subsubsection{Increased molecular gas mass}
Dynamical processes can drag atomic gas, initially at large radii, into the central regions of the galaxies where it will convert to molecular gas under the higher ambient pressures. \citet{kewley10} find significantly flatter gas-phase metallicity gradients in galaxies in close pairs,  compared to isolated galaxies due to the infall of metal-poor atomic gas from the outskirts as a result of the interaction \citep{barnes02, rupke10}. There is also evidence for an increase in the \hmol/\hi\ ratio as galaxies progress along a merger sequence \citep{mirabel89,lisenfeld19}. The cause of this is both an increase in the amount of molecular gas and the rapid consumption of the gas. For example, as systems move from the early to later stages of a merger sequence, their \hmol\ mass fraction doubles while their total cold ISM mass (\hi$+$\hmol) decreases, as a result of the \hi\ begin dragged inwards, converted to \hmol\ and then stars, or otherwise heated or removed from the system via outflows \citep{larson16, georgakakis00}.

\subsubsection{Shorter depletion timescales}
Extreme starbursts in the local Universe, observed in merging gas-rich systems, have \hmol\ depletion times $<10^8$~yr, an order of magnitude shorter than normal star-forming galaxies. The molecular gas in galaxies experiencing mergers is more centrally  concentrated and pushed to higher densities \citep{solomon97}, conditions that increase the efficiency of the gas-to-stars conversion.

In simulations, \citet{renaud14} find an increase in gas mass fraction in merger-induced tides, which pumps compressive turbulence into the ISM (as opposed to solenoidal turbulence). This produces an excess of dense gas, which leads to an increase in SFR per unit total molecular gas mass (diffuse$+$dense). The simulations of \citet{moreno21} also show that during the interaction of two galaxies, a central increase of the dense gas fraction is what mostly drives the SFR enhancement. This enhancement is modest in the initial phase of the interaction and highest around the time of coalescence. This is supported by the observations of \citet{pan18}, showing that star formation efficiency only increases significantly in galaxy-galaxy interactions at close distances ($<20$~kpc) and when the two galaxies are similar in mass. 

Several other studies also report that \tdep\ is shorter by a factor of 2-4 in galaxies showing signs of dynamical disturbances (minor mergers, galaxies with companions, galaxies with strong bars,...) \citep{young96, saintonge12, violino18}. While this enhancement in star formation efficiency is modest, these kinds of interactions are far more frequent than major mergers, making them important in an integrated sense. However, it should be noted that not all mergers and interactions lead to enhanced star formation; this has been observed and simulated \citep[e.g.][]{dimatteo07}, and linked to the specifics of the orbits of the two interacting galaxies and their relative properties, such as their mass ratio and gas masses.

\subsection{Below the main sequence}
\label{sec:belowMS}

What makes galaxies``fall off” the main sequence? All the so-called quenching mechanisms that have been proposed to explain the decrease of star formation in galaxies are related to the cold gas reservoirs, which have to be somehow depleted or prevented from forming new stars. Are the galaxies below the MS all gas-poor, and which cold gas phase is most affected? Can the decrease of SFR at fixed stellar mass be explained by reduced gas contents, or does it require a change in \tdep\ as well? Answering these questions ideally requires large, representative samples of galaxies below the MS with measurements of atomic and molecular gas, structural and star-formation properties. There are still very few studies of cold gas content of galaxies below the MS, partly due to a general perception that these systems are devoid of gas. While large integral-field spectroscopy (IFS) surveys are now underway, allowing detailed investigations of stellar populations, SFRs and chemical abundances on a spatially- (and spectroscopically-) resolved basis, the overlap between these and \hi\ and CO (even global) surveys is minimal, so that dedicated follow-ups of IFS data sets are required to measure cold gas content, especially below the MS. While this is a research area that will benefit tremendously from the increased sensitivity to small gas content afforded by the next-generation surveys with the SKA pathfinders and ALMA, several conclusions can be gleaned from existing work.\\

\begin{marginnote}[]
\entry{Quenching}{The process of halting star formation in a galaxy (e.g. via depletion of its cold gas reservoir). }
\entry{Green valley galaxies}{The population of galaxies with SFRs (or colors) intermediate between those of star-forming (blue) and quenched (red) galaxies.}
\end{marginnote}

\subsubsection{Decreased gas content} 
Both \fhi\ and \fgas\ decrease steadily below the main sequence. The gas content appears to decrease smoothly when galaxies move away from the MS at fixed stellar mass, without obvious discontinuities, until the sensitivity limits of the observations are reached (Figure \ref{fig:MS_gas}), and even further below once non-detections are taken into consideration (Figure \ref{fig:alongacross}b). For example, galaxies that are between 0.4 and 1.0 dex below the MS have offsets of 0.41 and 0.47 dex in \hi\ and \hmol\ gas fractions, respectively, compared to the averages on the MS, while these numbers fall to -1.10 and -1.21 dex for quenched systems (\deltams$<-1.0$ dex). 

Although there is significant scatter in \fhi\ and \fgas\ below the MS, there is no evidence for a significant population of “passive” galaxies with \hi\ or \hmol\ reservoirs that are comparable to those of MS galaxies. This is shown in Figure~\ref{fig:MS_gas}, as the galaxies below the main sequence, and especially those in the quiescent ``cloud" are for the most part non-detections in both CO and \hi, with just a handful of points that scatter towards gas fractions that are comparable to those of the MS. This contradicts recent claims of the existence of a large {\it population} of passive, massive discs harbouring \hi\ reservoirs as large as those of MS galaxies \citep[][but see \citealt{cortese20}]{zhang19}. This is not to say that there cannot be a small minority of unusual systems that are outliers in some of these scaling relations. Good examples of these are the so-called \hi-excess galaxies, which have unexpectedly large \hi\ reservoirs for their low SSFRs \citep{gereb16,gereb18}. Radio interferometric and optical follow-up observations indicate multiple possible reasons for their \hi\ excess, such as an external origin of the gas or unusually extended \hi\ disks with large specific angular momentum, although not all these systems are well understood at this point \citep{gereb18}. 

\begin{figure}[h]
\includegraphics[width=5in]{ 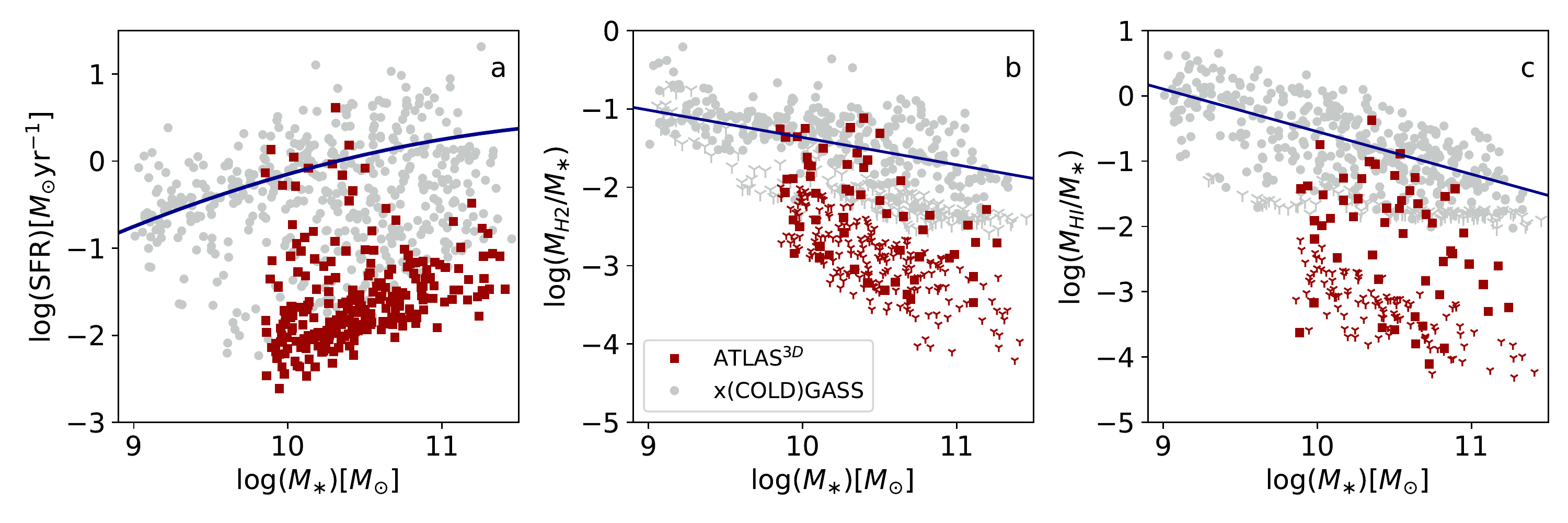}
\caption{Distribution in the SFR-\mstar\ plane (panel a) and comparison of \fgas\ (panel b) and \fhi\ (panel c) between the ATLAS$^{\rm 3D}$ (red) and xGASS/xCOLD GASS (gray) datasets. In both samples, gas detections and upper limits are indicated by small circles and downward arrows, respectively. Blue lines in all panels show the best fitting relation mass for main sequence galaxies ($|$\deltams$|<0.4$).}
\label{fig:atlas3d}
\end{figure}

A class of galaxies that is typically found below the MS is that of the early-types, i.e. ellipticals and lenticulars. Early studies reported low detection rates of cold gas in these galaxies \citep[e.g.,][]{gallagher75,knapp85,vandriel88,vandriel91},
reinforcing the perception that, although not always devoid of gas, these systems are generally gas-poor. Notably, the Infrared Astronomical Satellite (IRAS) detected large fractions of early-type galaxies at 60 and 100 $\mu$m \citep{jura86,knapp89}, implying the presence of a cold ISM in typical galaxies in this class and thus motivating more sensitive observations of their cold gas content \citep[e.g.,][]{sage89, wiklind89,roberts91}. These observations are now superseded by modern, more representative cold gas surveys of early-type galaxies selected from IFS surveys such as SAURON \citep{dezeeuw02}, ATLAS$^{\rm 3D}$ \citep{cappellari11} and MASSIVE \citep{ma14}. In particular the ATLAS$^{\rm 3D}$ \hi\ survey by \citet{serra12} detected 21 cm emission from 32\% of the 166 targeted galaxies, with the rate increasing to \about 40\% outside the Virgo cluster. Interestingly, a fraction of these systems were found to be as \hi-rich as spiral galaxies, but lacking the high column density gas typical of the latter. 
Detection rates of molecular gas were of order \about 25\% overall \citep{combes07,welch10,young11}, with a dependence on kinematic bulge fraction (i.e., systems characterised by higher stellar velocity dispersions at fixed mass are less likely to have detectable molecular gas) rather than local environment \citep{davis19}. Naturally, detection rates depend on the sensitivity of the observations (as well as sample selection) and might be misleading. Thus, we compare ATLAS$^{\rm 3D}$ \hi\ and \hmol\ measurements with xGASS and xCOLD GASS data in Figure~\ref{fig:atlas3d}. As can be seen, the populations of ATLAS$^{\rm 3D}$ early-type galaxies have significantly lower cold gas fractions compared to MS systems, with only a handful of exceptions, but those have SFRs putting them on the MS (see Figure~\ref{fig:atlas3d}a). In other words, statements that early-type galaxies are gas-poor systems and that they host significant cold gas reservoirs are not in contradiction, when {\it gas poor} is meant in comparison to galaxies with similar stellar mass on the MS and is not taken as synonym of devoid of gas. 

\subsubsection{Longer depletion timescales}
There is compelling evidence that changes in gas mass fraction alone cannot account for the full decrease of SSFR, but that it is accompanied by longer depletion times  \citep[e.g.][]{saintonge12,saintonge16,colombo20}. Longer depletion times below the MS might be caused by an increased stability of the molecular gas against fragmentation in bulge-dominated systems. Evidence for this comes for instance from the ATLAS$^{3D}$ early-type galaxies, which have longer \tdep\ than MS galaxies, due to increased shear in the inner parts of the galaxy (where the rotation curve is quickly rising and most \hmol\ is found; \citealt{davis14}) and/or increased velocity dispersion of the molecular gas \citep{dey19}. These longer depletion times might also be due to the fact that the CO(1-0) line traces both diffuse and dense molecular gas (see Section~\ref{sec:tdepvariations}). In other words, increasing diffuse-to-dense molecular gas mass ratios below the MS would lead to longer CO-based depletion times, even if the depletion time of the dense, star-forming gas was the same. As mentioned above, the \hi-emitting gas in these systems typically has lower column densities than in star-forming spirals, supporting the idea that the physical conditions of the cold gas reservoirs might be less conducive to star formation below the MS. 

Resolved studies that compare cold gas and star-formation properties of galaxies on and below the MS have the potential to shed light on the physical processes that drive the global trends discussed here. Several IFS studies have shown that the SFR density is suppressed at all radii in green valley galaxies, but especially in their central regions, suggestive of inside-out quenching \citep[e.g.,][]{belfiore18,medling18}.
In a recent work, \citet{brownson20} found that spatially resolved \hmol\ gas fraction ($\Sigma_{H_2}/\Sigma_\ast$) and SFE ($\Sigma_{SFR}/\Sigma_{H_2}$) are suppressed compared to typical star-forming regions, and that both contribute to reducing the resolved SSFR in seven green valley galaxies. In an interesting complementary approach, \citet{ellison21b} compared star-forming and {\it retired} spaxels of eight galaxies from the ALMaQUEST \citep{lin20} survey, and concluded that the retired spaxels are typically concentrated in the inner regions of galaxies and have significantly lower molecular gas fractions than the star-forming spaxels in the same galaxy, suggesting that quenching proceeds inside-out, at least for central galaxies.
Although based on small samples, such resolved studies are in qualitative agreement with the  findings based on global properties. The consensus is that a reduction of the molecular gas reservoir is a necessary but not sufficient condition to suppress star formation below the MS, as it must be accompanied by longer depletion times, as illustrated in Figure \ref{fig:alongacross}b.

\subsubsection{Multiple paths below the MS: the role of environment and AGN} 
The large scatter in both gas fractions and SFRs below the MS suggests that there is not a single physical process that is responsible for reducing the cold gas content and quenching the star formation in galaxies, and that there are likely to be multiple evolutionary paths through this region of parameter space. Indeed, we already know at least some of these different paths.

For satellite galaxies environmental effects are important, and there is now evidence that active stripping of the cold gas reservoirs is ubiquitous in both galaxy clusters and large groups -- although the overall picture across dark matter halo and stellar mass dimensions is a complex one, as recently reviewed by \citet{dawes9}. Environmental effects could also contribute to suppress star formation in the central galaxies of groups and clusters, by preventing their access to gas replenishment from large-scale filaments. 

For central isolated galaxies, {\it intrinsic} or {\it mass quenching} \citep{peng10}, i.e. halting of star formation caused by internal processes such as stellar or AGN feedback, is generally assumed to be more relevant. Super-winds and jets driven by starburst or AGN activity could impact star formation by directly removing part of the cold gas reservoir, or by injecting energy and turbulence into the gas and preventing cooling (negative feedback); locally, the same processes could also lead to enhanced star formation by compressing the gas (positive feedback). Numerical simulations and semi-analytic models of galaxy evolution have been invoking AGN feedback as a key process to quench the star formation in the most massive galaxies for over two decades \citep[e.g.,][]{kauffmann00,dimatteo05}. Indirect observational evidence for the impact of AGN activity on the ISM of galaxies comes for instance from the presence of large X-ray cavities in the hot ISM of the central galaxies of so-called cool core clusters \citep[see e.g. review by][]{fabian12}, as well as from the two large gamma-ray bubbles extending 50 degrees above and below the center of our own Milky Way \citep{su10,guo12}. Excitingly, fast, massive atomic and molecular gas outflows have also been observed in massive galaxies hosting AGN \citep[e.g.][]{morganti05,alatalo11,cicone14,morganti15}. However, the global cold gas reservoirs of galaxies harbouring an AGN do not seem to be significantly affected, as several studies reported similar molecular gas fractions and depletion times for systems with and without AGN located in the same region of the \ms\ plane \citep{saintonge17,rosario18,koss21}, a result that holds even for powerful quasars \citep{jarvis20}. The same result was found when comparing \hi\ gas mass fractions \citep[e.g.,][]{ho08,fabello11b,ellison19}.  Evidence therefore suggests that the biggest role played by AGN in the nearby Universe, at least when considering the large-scale properties of galaxies, is to keep halos hot. This throttles the accretion of new gas, pushing massive galaxies off the main sequence and preventing already quenched galaxies from resuming star formation activity \citep{best05}, but work continues to determine the impact of AGN activity on the cold ISM, especially on small scales.

\subsection{The mass-metallicity relation}

The scaling relation between stellar mass and gas-phase metallicity (i.e. the mass-metallicity relation, or MZR) provides additional clues to the importance of gas, and in particular the cold ISM, in regulating the growth of galaxies. The relation has been observed locally \citep[e.g.][]{lequeux79,tremonti04} but also up to $z>2$ \citep[e.g.][]{erb06}, and down in the regime of dwarf galaxies \citep[e.g.][]{lee06}. While stellar metallicities measure the integrated chemical evolution of galaxies over their entire star formation histories, gas-phase metallicity is a much more ``instantaneous" quantity, relying on the interplay of accretion of (mostly metal-poor) gas, star formation, and outflows of (mostly metal-enriched) material. Given its deep connection with all aspects of the baryon cycle, the MZR has huge value in constraining galaxy formation models, both analytic and numerical \citep[e.g.][]{edmunds90,dave12,lilly13,ma16}. 

The MZR has been extended to add a third parameter, in an effort to explain the correlated scatter of the relation.  For example at fixed \mstar, galaxies with higher SFRs are found to have lower gas-phase metallicities \citep{ellison08}. The SFR has been added to the MZR to form the \mstar-$Z$-SFR relation, referred to as the Fundamental Metallicity Relation \citep[FMR, e.g.][]{mannucci10}.   There is however evidence that gas mass is the more fundamental parameter associated with the scatter of the MZR, with a strong observed anti-correlation found between \mhi\ and gas-phase metallicity at fixed stellar mass \citep{bothwell13FMR,laralopez13,hughes13FMR,brown18}. \citet{bothwell16} also report that the scatter of the MZR correlates with \mh. These studies all find that the scatter of the MZR depends more strongly on gas mass than on SFR, which would make the FMR a consequence of this more fundamental relation between stellar mass, metallicity and gas mass (via the link between gas and star formation). 

There are important practical implications for this result. For example, the suggestion of a redshift-independent relation between metallicity and the stellar mass-to-gas mass ratio with small intrinsic scatter \citep{zahid14} might provide an alternative route to predicting gas masses from optical spectroscopy (see Sec. \ref{sec:indirecttracers}). The MZR also offers opportunities to constrain important elements of the baryon cycle that are otherwise difficult to directly observe For example, strong outliers from the MZR with low metallicities given their mass, could represent objects having recently undergone an accretion event \citep[e.g.][]{peeples09}.

\subsection{The multiple scales of star formation: from halos to molecular clouds}
\label{sec:tdepvariations}

In the sections above we have highlighted how the star formation activity and metallicity of galaxies are tightly linked to their cold ISM properties. The \hi\ and \hmol\ contents of galaxies vary smoothly as a function of both \mstar\ and SFR, as does \tdep, and as a result the position of galaxies in the SFR-\mstar\ plane can be explained by (1) the total mass of their cold ISM reservoirs ($M_{gas}=$\mhi$+$\mh), (2) the fraction of this total gas that is in molecular form (\mh$/$\mgas), and (3) the timescale associated with the conversion of this molecular gas into stars ($t_{dep,H2}$): 
\begin{align}
{\rm SFR} &= \frac{M_{gas} \cdot \frac{M_{H2}}{M_{gas}}}{t_{dep,H2}}, 
\end{align}
 \label{eq:sfrmodel}
where each term on the right-hand-side is a function of \mstar\ and \deltams. As illustrated in Figures \ref{fig:MS_gas} and \ref{fig:alongacross}, the relative importance of the different terms on the right hand side of Eq. 8 varies for different galaxy sub-populations. For example, the shape and scatter of the MS is mostly driven by \mh, high SFRs in starburst galaxies are the result of both an increase in \fgas\ (though a more efficient atomic-to-molecular conversion) and shorter \tdep, while low SFRs in quenched galaxies are mostly driven by low \mgas.

\begin{figure}[h]
\includegraphics[width=5in]{ 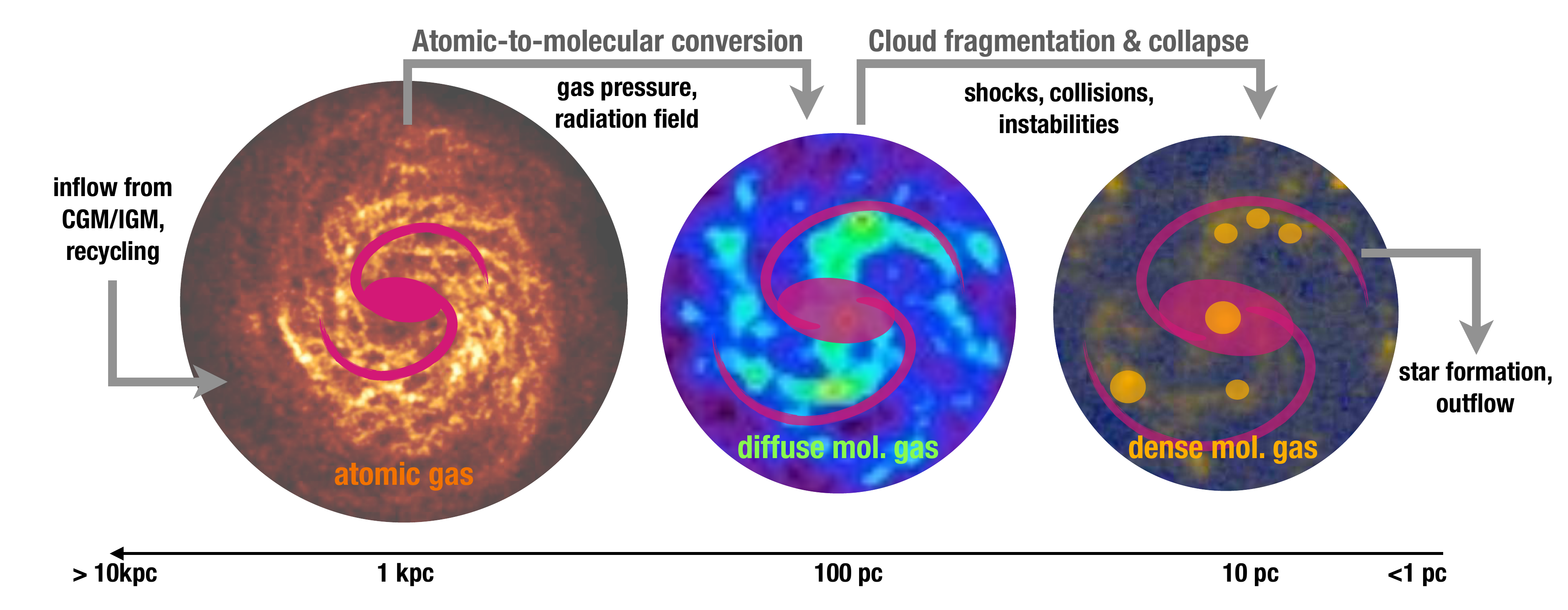}
\caption{Illustration of the connections between the cold ISM and the other components of the baryon cycle, focusing on the path of gas from the large scale environment all the way to star formation. We highlight the range of scales characteristic of each step of the process, and the physical mechanisms at play.  The \hi-to-\hmol\ transition mostly depends on midplane gas pressure \citep[e.g.][]{blitz04,leroy08}, with contributions from the radiation field \citep[e.g.][]{elmegreen93}, while further collapse and fragmentation of GMCs is linked to shocks (from SNe, for example), turbulence, and magnetic fields.}
\label{fig:summary}
\end{figure}

Interestingly, the terms on the right hand side of Equation 8 are each regulated by physical mechanisms operating on very different scales, placing the cold ISM at the crossroads. This is represented schematically in Figure \ref{fig:summary}. For example, both observations and simulations suggest a deep connection between $M_{gas}$ and the angular momentum and formation histories of the dark matter halos in which the galaxies reside \citep[e.g.][]{boissier01,obreschkow16,guo17,lutz18,obuljen19,mancera21}. This is an active area of research that will benefit from the sensitivity of the next generation of observations, especially to push the study of the galaxy-halo connection to lower mass galaxies. Such a paradigm may enable the use of \hi\ observations to (indirectly) probe the very largest scales of the galactic ecosystem, the dark matter halos, and their histories. 

The second term of Equation 8, the molecular ratio (\mh/$M_{gas}$), is on the other hand controlled by the conditions within galactic discs, in particular the gas density, strength of the UV field, and metallicity  \citep[e.g.][]{elmegreen93,gk11}. For example, many observations show that the atomic-to-molecular ratio correlates with the midplane hydrostatic gas pressure \citep[e.g.][]{blitz04,leroy08}. A very significant development over the past decade has been the realisation that the properties of GMCs themselves also  depend on the large-scale properties of the galactic disc in which they are found \citep[e.g.][]{hughes13,sun20}, responding to the pressure of their environment \citep[e.g.][]{field11,meidt18}. This is a significant shift from the previous picture of ``universal" GMCs with a constant depletion time \citep[e.g.][]{leroy08,bolatto08}, spurred by the observation of the systematic trends in \tdep\ across the local galaxy population (see Section \ref{sec:tdep}), and the ability to map CO at resolution of tens of parsec in galaxies spanning a wider range of properties \citep[e.g.][]{schinnerer13,schruba19}.

Once it is recognised that the properties of GMCs can vary within galaxies, and from galaxy-to-galaxy, the observed variations in \tdep\ can be partly explained by the specifics of the molecular gas tracer employed, the CO(1-0) emission line. With a critical density of $\sim10^3$~cm$^{-3}$, a significant (but variable) fraction of the total CO(1-0) emission traces diffuse gas, while star formation mostly takes place in the densest regions of molecular clouds \citep{heiderman10}. In the Milky Way, for example, 25\% of the molecular gas is diffuse (defined as gas detected in $^{12}$CO but not in $^{13}$CO), in fact only about 15\% of the molecular gas mass traced by $^{12}$CO is in observed molecular cloud complexes \citep{romanduval16}. 

Variations due to changes in the dense-to-diffuse molecular gas ratio can be captured by explicitly distinguishing between the total \mh\ (as traced by e.g. CO(1-0)) and the dense star-forming gas in Equation \ref{eq:sfrmodel}:
\begin{align}
{\rm SFR} &= \frac{M_{gas} \cdot \frac{M_{H2}}{M_{gas}} \cdot \frac{M_{dense}}{M_{H2}} }{t_{dep,dense}}, 
\end{align}
 \label{eq:sfrmodel_2}
where $M_{dense}$ represents the mass of dense molecular gas, say with density in excess of $\sim10^5$~cm$^{-3}$, and $t_{dep,dense}=M_{dense}/{\rm SFR}$. Efforts are ongoing to determine how $M_{dense}/M_{H2}$ and $t_{dep,dense}$ might vary systematically both within galaxies and from galaxy-to-galaxy, so far with a focus on using HCN and HCO$^+$. With higher effective critical densities even for their low-$J$ lines, these molecules probe the star-forming gas more directly. \citet{gao04} indeed reported a linear relation between $L_{\rm HCN}$ and $L_{\rm IR}$ (the latter a proxy for SFR) and argue that SFR depends on the amount of dense molecular gas traced by HCN emission, rather than the total molecular gas mass traced by CO emission. 

\begin{figure}[h]
\includegraphics[width=5in]{ 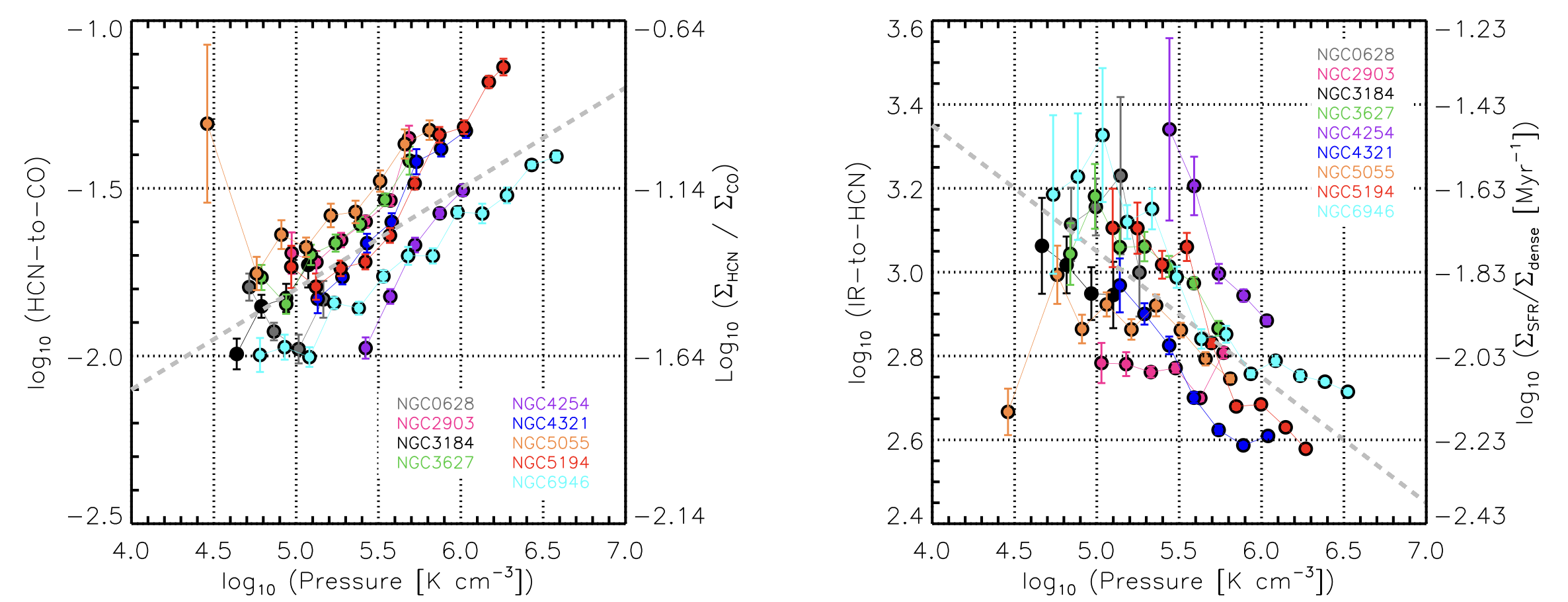}
\caption{Variations in the dense-to-diffuse molecular gas ratio as measured by the HCN/CO luminosity ratio (left) and in the star formation rate per unit dense gas mass (right) as a function of local pressure in 9 galaxies from the EMPIRE survey. Figure reproduced with permission from \citet{jimenez19}.}
\label{fig:EMPIRE}
\end{figure}

The EMPIRE program \citep{bigiel16} was designed to look for systematic variations in the dense gas ratio and $t_{dep,dense}$ by mapping HCN(1-0), HCO$^+$(1-0), HNC(1-0) and CO isotopologues at \about 1-2 kpc resolution in nine nearby massive spiral galaxies. The observations reveal that the HCN/CO ratio varies systematically within galactic discs and is highest in high pressure environments \citep{usero15}, as shown in Figure \ref{fig:EMPIRE} (left). Interestingly, even after accounting for the different conditions within the discs of each galaxy, galaxy-to-galaxy variations in the HCN/CO ratio remain, correlating with global galaxy properties such as \must\ and the \hi-to-\hmol\ ratio \citep{jimenez19}.

There are also systematic variations in star formation efficiency with global galaxy properties, even when only accounting for the HCN-traced dense gas (Figure \ref{fig:EMPIRE}, right). This result is particularly interesting, as it goes against simple expectations, even showing that the dense gas star formation efficiency is {\em lower} in high pressure regions. This could be telling us that HCN does not perfectly trace the dense star-forming gas, and that with a different tracer with an even higher critical density we would reach the point of a constant $t_{dep,dense}$ as proposed in certain theories of star formation \citep[e.g.][]{evans14}. Alternatively, Figure \ref{fig:EMPIRE} (right) could instead be telling us that our expectation of a universal dense gas depletion time above a certain gas density threshold is not correct, which would argue in favour of turbulence-driven theories of star formation \citep[e.g.][]{federrath12}. This is challenging work with current facilities. For example, HCN(1-0) is \about15-30 times fainter than CO(1-0) emission \citep{gao04}, limiting studies to modest samples of mostly main-sequence and starburst galaxies. Further studies will however be key to quantify scaling relations and test star formation theories, as discussed in Section \ref{sec:Outlook} below. 

\section{CLOSING THOUGHTS AND OUTLOOK}
\label{sec:Outlook}

The past couple of decades have seen tremendous progress in the study of the cold gas in galaxies, with the undertaking of several large, dedicated observing campaigns that have measured the atomic and molecular hydrogen content of sizeable samples of galaxies on global and local scales, and placed these in the context of their optical and star-formation properties, as well as environments. In this review we have focused on global \hi\ and \hmol\ scaling relations in the local Universe ($z<0.05$), and discussed the link between cold gas content and star formation of galaxies across the SFR-\mstar\ plane.

Overall, the empirical global cold gas scaling relations seem well established, at least for galaxies with stellar masses $M_{\ast}>10^9 M_{\odot}$, for which the systematic variation of atomic and molecular gas fractions has been quantified as a function of stellar mass and other properties (e.g. Figure~\ref{fig:scaling_mass}). For galaxies with stellar masses below this threshold, global \hi\ measurements do exist \citep[e.g.,][]{geha06,bradford15}, but systematic CO measurements are still rare and dominated by upper limits \citep[e.g.,][]{lisenfeld11,cicone17}. These have been compiled and homogenised by \citet{calette18} and 
\citet{hunt20} to extend the cold gas fraction-stellar mass relations down to $M_{\ast} \sim 10^7 M_{\odot}$. There is tantalizing evidence for a break at $M_{\ast} \sim 10^9 M_{\odot}$ in the \hi\ scaling relation \citep[see also][based on ALFALFA]{maddox15}, but more statistics is needed to confirm these results. 

Furthermore, much remains to be learned about the physical drivers of these relations and how the global gas properties of galaxies are linked to their molecular clouds and dense core distributions, and on the other side to the large scales via the surrounding CGM (which extends roughly from the stellar disk to the virial radius of a galaxy), where inflows and outflows contribute to replenish or reducing the cold gas in the ISM. In what follows, we briefly reflect on how our view of galaxies and their gas reservoirs has changed in the past decade and highlight a few areas where future work will allow us to make progress in this field.\\

\noindent{\bf A changing view of the gas reservoirs of galaxies} 
Galaxies are no longer viewed as isolated entities, but as systems embedded in a cosmic web whose evolution is regulated by exchanges of gas with the surrounding environment. This gas flows from the Intergalactic Medium (IGM) through the CGM and into the ISM, where it is processed into stars, and then partly returned to the CGM/IGM via outflows from supernova explosions, stellar winds, or nuclear activity. Through this cycle, properties such as temperature and density of the gas change by orders of magnitude, from the tenuous, hot ionised IGM to the dense, cold neutral cores of GMCs. Focussing on the ISM (the topic of this review), various tracers are used to measure the global gas content in its different physical states, each with its associated depletion time, which is the time needed to consume the total amount of gas in the given phase via star formation, assuming constant SFR and no recycling. The fact that depletion times increase when moving from dense to diffuse gas phases means that they somehow encapsulate information on how close the given gas phase is to the star-formation process -- in a way, akin to clocks that count down to star formation. Although the inverse of a depletion time is usually referred to as star-formation {\it efficiency}, this quantity tells us very little about how star formation proceeds at the scale of GMCs, where new stars are actually formed (efficiency should also be a dimensionless quantity). In this context, all these gas phases are {\it transient}, i.e. way stations toward star formation, rather than ``reservoirs'' -- although it is convenient to picture these as such when depletion times are of order of a Gyr or more. In order for a galaxy to form stars, its dense gas phase needs to be constantly re-supplied, ultimately from the ionised IGM, which is the reservoir that fuels star formation at all epochs and is replenished by outflows.\\

\noindent{\bf Connecting ISM and CGM.}
The interface between the ISM and the CGM, which is where gas is exchanged and recycled, is of crucial importance for our understanding of galaxy evolution, but we lack an empirically motivated model for how these gas flows proceed \citep[see however][]{gaspari13}, and how they may vary across the galaxy population. This is an important topic beyond the scope of this review, but it is nonetheless worth identifying areas where future cold gas observations might help. Because of its low density, the CGM is best studied via quasar absorption spectroscopy, a field that has benefited enormously from the availability of the Cosmic Origins Spectrograph on the Hubble Space Telescope \citep{tumlinson17}. Interestingly, studies of galaxy-quasar pairs (where galaxies happen to be near the line-of-sights to background quasars) are helping to shed light on the connection between ISM and CGM, by investigating how the strengths of ultraviolet absorption features tracing the CGM vary with impact parameter and galaxy properties such as \hi\ content and SFR \citep[e.g.,][]{borthakur13,borthakur15,borthakur16}. Complementary to this approach, surveys probing multiple quasar sightlines around the same nearby galaxy, such as project AMIGA \citep[Absorption Maps In the Gas of Andromeda;][]{lehner20}, can better illustrate how the structure and properties of the CGM vary with galactocentric distance, perhaps in relation to a more direct impact of galactic feedback in the inner regions of the CGM. These observational efforts are accompanied by significant efforts to simulate the CGM and its connection with the ISM at high resolution \citep[e.g.][]{peeples19,vandevoort19}. In the future, larger samples of galaxies with resolved and sensitive \hi\ and \hmol\ measurements, probing the low column density outer regions of galaxy disks, will allow us to extend these studies and explore the ISM/CGM connection in greater detail. \\ 

\noindent {\bf Connecting global gas and local star formation.}
Further down the line of the baryon cycle, tremendous progress has also been made over the past decade to connect the process of star formation to global galaxy properties. There is nonetheless still a striking disconnect between our ability to investigate the link between gas and star formation at the scale of individual cores in the Milky Way, and studies of external galaxies, where we are not yet able to probe the spatial scales at which the physics of star formation is important. A key challenge for galaxy evolution studies is to connect these widely different scales across the galaxy population. Surveys of cold gas in nearby galaxies at kpc resolution (e.g., THINGS, HERACLES) have been invaluable to start bridging this gap, and high-resolution studies of individual galaxies at tens of pc resolution \citep[e.g. PAWS,][]{schinnerer13} are pushing this even further. The PHANGS \citep{leroy21} survey has carried out arcsecond CO(2-1) imaging of 70 nearby star-forming galaxies with ALMA, reaching the size of typical GMCs (\about 100 pc resolution). These efforts are providing key empirical constraints on the physical link between star formation and gas near the ``cloud" scale and the galaxy-scale environment  \citep[e.g.,][]{hughes13,sun18,schruba19,chevance20,sun20,liu21}.  Further progress will require even larger samples, extending the parameter space especially towards low mass galaxies and below the main sequence. The synergy between high-resolution, sensitive cold gas studies and IFS  surveys will be critical to advance this field.

The connection between gas and star formation will further benefit from exploiting a larger range of cold ISM tracers, in particular molecular lines with high critical densities such as HCN, HCO$^+$, CS, or even higher-$J$ CO lines. To expand on the current state of the art (e.g. the EMPIRE \citep{jimenez19} and MALATANG \citep{MALATANG1} surveys), we will need larger samples, especially including galaxies below the star-forming sequence where CO-based depletion times are longer. Even if spatial resolution has to be sacrificed to assemble such sample (or if spectral stacking has to be used), there would be much to be gained from dense gas versions of the scaling relations presented in this review. Comparing the shape and scatter of the scaling relations from diffuse and dense gas tracers will shed light on the role of secular and environmental processes regulating the global star formation activity of galaxies. Mapping the dense gas at resolutions of tens of pc in galaxies displaying a wide range of internal conditions will also be crucial to put to the test various models for star formation, for example confronting density threshold and turbulence-driven models. \\

In conclusion, much progress is being made towards our understanding of the gas-star formation cycle via multi-wavelength, multi-scale investigations. 
For {\em global} gas mass measurements, further breakthroughs will require large representative samples that extend the current parameter space coverage, especially pushing into the regime of low mass (dwarf) galaxies (\mstar$<10^9$\msun). The Five-hundred-meter Aperture Spherical radio Telescope \citep[FAST;][]{FAST} will have a role to play in these efforts for \hi\ studies, complementary to that of wide-area, blind \hi\ surveys such as WALLABY \citep{wallaby} on the Australian Square Kilometre Array Pathfinder \citep[ASKAP;][]{johnston08,hotan21}. Similarly, the next generation of (sub)millimeter single-dish telescopes (for example the AtLAST concept\footnote{https://www.atlast.uio.no}) will allow for orders of magnitude increase on the current molecular gas sample sizes. However, extending large molecular gas surveys into the regime of low mass (and therefore low metallicity) galaxies will first require further calibration of our tracers of the molecular ISM, as both CO- and dust-based methods present challenges under such conditions. A next generation far-infrared telescope, such as the Origins Space Telescope\footnote{https://asd.gsfc.nasa.gov/firs/}, would significantly help in these efforts.

Our view of the link between gas, star formation, and global galaxy properties will be revolutionised by {\em resolving} atomic and molecular gas in large samples containing not only star-forming galaxies, but spanning the full range of stellar masses, environments, size and angular momentum, and including systems lying significantly off the main sequence (either above or below). 
This has been particularly challenging for cold gas surveys so far, traditionally limited by a trade-off between spatial resolution and number statistics, and for \hi\ emission studies also restricted to the very local Universe for individual detections \citep[i.e., $z<0.4$;][]{highz,fernandez16}. But there is much to look forward to in the coming decade(s), thanks to the powerful synergy between IFS instruments on 10 meter-class (and higher) telescopes on one hand, and ALMA, the Square Kilometre Array \citep[SKA;][]{SKA} and its precursor facilities (e.g., ASKAP and the South African Meer-Karoo Array Telescope, MeerKAT, \citealt{jonas16}) on the other. While the former will measure star-formation, metallicity and stellar properties in exquisite detail, the latter will probe the cold gas component with unprecedented sensitivity and resolution, also pushing \hi\ detections to redshift 1 and beyond. \\

\begin{summary}[SUMMARY POINTS]
\begin{enumerate}
\item The cold ISM in the nearby Universe accounts for $\sim1.3\%$ of all the baryons, and 35\% of the total mass density of stars. Massive galaxies (\mstar$>10^{10}$\msun) account of $\sim65\%$ of all this molecular gas, but only $\sim40\%$ of the atomic gas. 
\item The atomic and molecular gas mass fractions vary systematically across the galaxy population. There is a sharp decline in both atomic and molecular gas contents above \mstar$\gtrsim 10^{10.3}$\msun. Below this threshold, the \mh-to-\mhi\ ratio decreases steadily; low mass galaxies have extended atomic gas reservoirs that are largely decoupled from the star formation process. 
\item The depletion timescale for \hi\ is on average \about3 Gyr, but with very large galaxy-to-galaxy variations which dominate over any systematic trends. On the other hand, \tdep\ varies systematically, in particular as a function of SSFR (or offset from the main sequence). These variations are likely caused by changes in the dense-to-diffuse molecular gas ratio.  
\item The shape of the star formation main sequence is set mostly by the dependence of \mh\ on \mstar\, and to second order by a slowly increasing \tdep\ with mass.  Star-forming galaxies just above (below) the main sequence have higher (lower) molecular gas fractions, suggesting that the scatter of the main sequence is gas-driven and that main sequence offsets persist on timescales of $\sim$Gyr. 
\item Starburst galaxies reach their high levels of star formation due to a combination of increased molecular gas content (via a more efficient atomic-to-molecular gas conversion) and decreased global depletion times. 
\item Below the MS, galaxies are mostly characterised by lower cold gas reservoirs, with a second-order contribution from longer depletion times (again, likely related to variations in the dense-to-diffuse molecular gas ratio). There is no evidence for a significant population of quiescent but gas-rich galaxies. 
\item Overall, the position of galaxies in the SFR-\mstar\ plane can be explained by their gas contents and the global efficiency of the gas-to-star conversion process. This highlights the importance of phenomena on a huge range of physical scales in driving galactic evolution: from cosmic web-regulated gas accretion, to cloud-scale star formation. 
\end{enumerate}
\end{summary}

\begin{issues}[FUTURE ISSUES]
\begin{enumerate}
\item Extending current efforts to establish cold ISM scaling relations into the regime of low mass galaxies (\mstar$<10^{9}$\msun) beyond current studies is needed to better understand the baryon cycle and processes such as cosmic reionisation and the galaxy-halo connection. This will require the sensitivity of new facilities, and significant work in understanding the systematics of the cold ISM tracers in extreme environments (e.g. very low metallicities, or very strong UV radiation fields). 
\item Mapping of atomic and molecular gas (using both diffuse and dense gas tracers) at high resolution (tens of pc) across a broad range of galaxy properties and well into the quiescent population will be required to uncover the multi-scale nature of the star formation process. 
\item Exploring the connection between the cold ISM and the other gaseous phases of the galactic environment (in particular the CGM and IGM) as well as their dark matter halos will be key in understanding what regulates the availability of gas and therefore, down the line and many orders of magnitudes in physical scales, the star formation activity of galaxies. 
\end{enumerate}
\end{issues}

\section*{DISCLOSURE STATEMENT}
The authors are not aware of any affiliations, memberships, funding, or financial holdings that
might be perceived as affecting the objectivity of this review. 

\section*{ACKNOWLEDGMENTS}
We thank Luca Cortese for useful discussions and for commenting on an early draft of this review, as well as Romeel Dav\'e, Tim Davis, Sara Ellison, Luis Ho, Rob Kennicutt, Lihwai Lin, David Rosario and Stijn Wuyts for their input. We are grateful to Caroline Bertemes, Chiara Circosta, Robert Feldmann and Steven Janowiecki for making data and code available. 

Parts of this research were conducted by the Australian Research Council
Centre of Excellence for All Sky Astrophysics in 3 Dimensions (ASTRO 3D),
through project number CE170100013. AS acknowledges support from the Royal Society. 



\end{document}